\documentclass[twocolumn]{aastex631}

\begin{document}

\title{Determining the Timescale over Which Stellar Feedback Drives Turbulence in the ISM: A Study of four Nearby Dwarf Irregular Galaxies}

\author[0000-0001-5368-3632]{Laura Congreve Hunter}
\affiliation{Department of Astronomy, Indiana University, 727 East 3rd Street, Bloomington, IN 47405, USA}

\author{Liese van Zee}
\affiliation{Department of Astronomy, Indiana University, 727 East 3rd Street, Bloomington, IN 47405, USA}

\author[0000-0001-5538-2614]{Kristen B. W. McQuinn}
\affiliation{Rutgers University, Department of Physics and Astronomy, 136 Frelinghuysen Road, Piscataway, NJ 08854, USA}

\author[0000-0002-9426-7456]{Ray Garner, III}
\affiliation{Department of Astronomy, Case Western Reserve University, 10900 Euclid Avenue, Cleveland, OH 44106, USA}

\author{Andrew E. Dolphin}
\affiliation{Raytheon Company, 1151 E. Hermans Road, Tucson, AZ 85756, USA}
\affiliation{University of Arizona, Steward Observatory, 933 North Cherry Avenue, Tucson, AZ 85721, USA}

\begin{abstract}
    Stellar feedback is fundamental to the modeling of galaxy evolution as it drives turbulence and outflows in galaxies. Understanding the timescales involved are critical for constraining the impact of stellar feedback on the interstellar medium (ISM).  We analyzed the resolved star formation histories along with the spatial distribution and kinematics of the atomic and ionized gas of four nearby star-forming dwarf galaxies (NGC 4068, NGC 4163, NGC 6789, UGC 9128) to determine the timescales over which stellar feedback drives turbulence. The four galaxies are within 5 Mpc and have a range of  properties including current star formation rates of 0.0005 to 0.01 M$_{\odot}$ yr$^{-1}$, log(M$_*$/M$_{\odot}$) between 7.2 and 8.2, and log(M$_{HI}$/M$_\odot$) between 7.2 and 8.3. Their Color-Magnitude Diagram (CMD) derived star formation histories over the past 500 Myrs were compared to their atomic and ionized gas velocity dispersion and   HI energy surface densities as indicators of turbulence. The Spearman's rank correlation coefficient was used to identify any correlations between their current turbulence and their past star formation activity on local scales ($\sim$400 pc). The strongest correlation found was between the   HI turbulence measures and the star formation rate 100-200 Myrs ago.  This suggests a coupling between the star formation activity and atomic gas on this timescale.  No strong correlation between the ionized gas velocity dispersion and the star formation activity between 5-500 Myrs ago was found.  The sample and analysis are the foundation of a larger program aimed at understanding the timescales over which stellar feedback drives turbulence.
\end{abstract}

\section{Introduction}

Star formation activity is thought to drive turbulence in the interstellar medium (ISM) through ionizing radiation, stellar winds, and supernovae (SNe) (e.g., \citealt{Spitzer78,Elmegreen04,maclow04}). As a necessary component in understanding galaxy evolution, stellar feedback and turbulence are invoked to regulate star formation and star formation efficiencies  (e.g., \citealt{OstrikerShetty11}), to drive the loss of metals via outflows and explain the observed galaxy mass-metallicity relationship (e.g., \citealt{Tremonti04,Brooks07,Christensen18}), and are proposed as a solution to the core-cusp dispute over dwarf galaxies' dark matter distributions (see \citealt{Bullock17} for a review).  Observational studies have found a correlation between current star formation and the H$\alpha$ velocity dispersion ($\sigma_{H\alpha}$) \citep{Moiseev15,Yu19}, as well as a correlation between the current star formation activity and HI turbulence at high star formation rate (SFR) surface density (e.g, \citealt{Joung09,Tamburro09,Stilp13c}).  However, a correlation between the current star formation activity and HI turbulence is not seen in regions of low SFR surface density such as the outer regions of spirals and in dwarf galaxies (e.g, \citealt{vanzee99,Tamburro09}).

From recent theoretical work, there are suggestions that the impact of recent star formation activity may not be immediately observable as turbulence.  These models observe a time delay between star formation activity and stellar feedback driven turbulence in the ISM (e.g., \citealt{Braun12}).  From the FIRE \citep{Hopkins14} and FIRE-2 \citep{Hopkins18} simulations, enhancement of the ionized gas velocity dispersion and instantaneous SFR may be asynchronous on quite short timescales (less than tens of Myrs; \citealt{Hung19}), while an increase in the atomic gas velocity dispersion may be more closely correlated with increased star formation activity on longer timescales ($\sim$ 100 Myrs; \citealt{Orr2020}).  To identify such a correlation between the current ISM turbulence and the past star formation activity, the use of time resolved star formation histories (SFHs) is required.  

Most previous observational work studying the relationship between stellar feedback and the turbulence in the ISM has focused on integrated light measurements to determine SFRs (e.g, \citealt{Zhou17,Hunter21}). While observationally less expensive, SFRs based on integrated light measurements are limited to set timescales ($<$10 Myrs for H$\alpha$ and $<$100 Myr for the far UV: see \citealt{Kennicutt12} and references therein), and they are not sensitive to time variability in the SFR. However, the time-variability of galaxies' SFHs has been well established (e.g, \citealt{Dolphin05, McQuinn10a,McQuinn10b, Weisz11,Weisz14}). Thus, time resolved SFHs are needed to analyze the impact of star formation on the ISM over time.  Resolved stellar populations provide a means to measure the SFR as a function of time and study time-variable star formation, stellar feedback, and galaxy evolution \citep{Dalcanton09,McQuinn10a,Stilp13b}. Color-magnitude diagrams (CMDs) can reconstruct the SFH using stellar evolution isochrones and CMD fitting techniques (e.g, \citealt{Tolstoy96,Dolphin97,Holtzman99,Harris01,Aparicio09}).  By comparing the current HI turbulence and CMD derived SFHs for a sample of 18 dwarf galaxies, \cite{Stilp13b} found a correlation between the global HI energy surface density ($\Sigma_{HI}$) and the star formation activity 30-40 Myrs ago.

In this paper, we focus on the correlation timescale on local scales. We explain our methodologies for connecting recent star formation with turbulence in multiple phases of the ISM in low-mass galaxies in 400 parsec regions.  The analysis focuses on a subsample of four galaxies (NGC 4068, NGC 4163, NGC 6789, and UGC 9128) from a larger sample of low-mass galaxies, as a demonstration of the analysis strategies. Section \ref{obs} discusses the data used in this study from the Very Large Array (VLA\footnote{The VLA is operated by the NRAO, which is a facility of the National Science Foundation operated under cooperative agreement by Associated Universities, Inc.}), \textit{Hubble Space Telescope (HST)}, and WIYN \footnote{The WIYN Observatory is a joint facility of the NSF's National Optical-Infrared Astronomy Research Laboratory, Indiana University, the University of Wisconsin-Madison, Pennsylvania State University, the University of Missouri, the University of California-Irvine, and Purdue University. } 3.5m telescope, and Section \ref{regions} outlines how we measure the turbulence of the atomic and ionized gas and determine SFHs.  Section \ref{results} presents our initial results on the timescales over which stellar feedback drives turbulence. Section \ref{conclusion} summarizes the initial results and outlines the upcoming larger project.

\section{Observational Data}\label{obs}

For our initial study on the impact of stellar feedback on turbulence in the ISM, multi-wavelength observations have been acquired for four nearby low-mass galaxies (NGC 4068, NGC 4163, NGC 6789, and UGC 9128; see Table \ref{table:galax}).   All four of these galaxies have previously had their global SFHs determined as part of \cite{McQuinn10a} and are part of STARBIRDS \citep{McQuinn15a}.  These systems were selected as a pilot study to test our methodology for connecting star formation timescales with gas kinematics as they are representative of the larger sample. NGC 4068, at almost 4.4 Mpc, and UGC 9128, as one of the lower surface brightness galaxies, are good tests of our abilities to accurately recover spatially resolved SFHs. In addition, UGC 9128 and NGC 6789 have physical sizes that result in small numbers of regions and are useful for determining how to best partition the galaxies. 

A combination of new and archival Very Large Array (VLA) radio synthesis observations (new observations listed in Table \ref{table:hiobs}) of the neutral hydrogen 21 cm emission line are used to determine the atomic gas surface density and velocity dispersions (see Table \ref{table:cubes}). Archival F555W, F606W, and F814W \textit{Hubble Space Telescope (HST)} observations of resolved stars were used to create CMDs, from which we derive SFHs (see Table \ref{table:HST_obs}).  Spectroscopic Integral Field Unit (IFU) observations from SparsePak on the WIYN 3.5m telescope provide the ionized gas kinematics (Table \ref{table:optobs}).  

\begin{table*}
    \scriptsize
    \caption{Galaxy Sample and Optical Properties}
    \centering 
        \begin{tabular}{c c c c c c c c c c c c c c } 
        \hline
        \hline 
        Galaxy & RA & Dec & Dist & m$_B$ & A$_B$ & M$_B$ & D$_{25}$ & B/A  & M$_*$ & log(H$\alpha$) & log(SFR)\\
         & J2000 & J2000 & Mpc & mag & mag & mag & arcsec &  & log(M$_\odot$) & erg s$^{-1}$cm$^{-2}$  & M$_\odot$ yr$^{-1}$ \\
         (1) & (2) & (3) & (4) & (5) & (6) & (7) & (8) & (9) & (10) & (11) &(12) \\
         [0.5ex] 
        \hline 
        NGC 4068 & 12:04:03 & 52:35:29 & 4.38$\pm$0.04 & 13.10 & 0.09 & -15.20$\pm$0.02 & 167.4 & 0.54 & 8.34$\pm$0.07 & -12.08$\pm$0.05& -1.98$\pm$0.05 \\
        NGC 4163 & 12:12:09 & 36:10:10 & 2.88$\pm$0.04 & 13.46 & 0.09 & -13.92$\pm$0.03 & 116.0 & 0.67 & 7.99$\pm$0.12 & -12.67$\pm$0.05& -2.94$\pm$0.05 \\
        NGC 6789 & 19:16:42 & 63:58:16 & 3.55$\pm$.007 & 14.02 & 0.3 & -14.03$\pm$0.03 & 84.8 & 0.85 & 8$.0\pm$0.13 & -12.88$\pm$0.06& -2.97$\pm$0.06 \\
        UGC 9128 & 14:15:57 & 23:03:22 & 2.21$\pm$0.07 & 14.39 & 0.10 & -12.43$\pm$0.07 & 92.2 & 0.67 & 7.11$\pm$0.07 & -13.79$\pm$0.11& -4.29$\pm$0.11 \\[1ex] 
        \hline 
    \end{tabular} 
    \tablecomments{\scriptsize Column (4) Distances from CMD fitting for UGC 9128 and NGC 4163 are from \cite{Dalcanton09} and NGC 4068 and NGC 6789 distances are from \cite{Tully13}. Column (5) from WIYN 0.9m imaging taken on Sept. 28 2007, Column (6) Galactic extinction values from \cite{Schlegel98}, Columns (7-9) from WIYN 0.9m imaging taken on Sept. 28 2007, Column (10) Stellar masses from \cite{McQuinn19} except NGC6789 which is from \cite{McQuinn10b} Columns (11) from \cite{McQuinn19} except NGC 6789 which the H$\alpha$ flux and SFR based off WIYN 0.9m imaging taken on Sept. 15 2007 Column (12) H$\alpha$ SFR based off equations presented in \cite{Kennicutt12}}
    \label{table:galax} 
\end{table*}

\subsection{VLA Observations}

\begin{figure*}[!th]
    \centering
    \includegraphics[width=.9\textwidth]{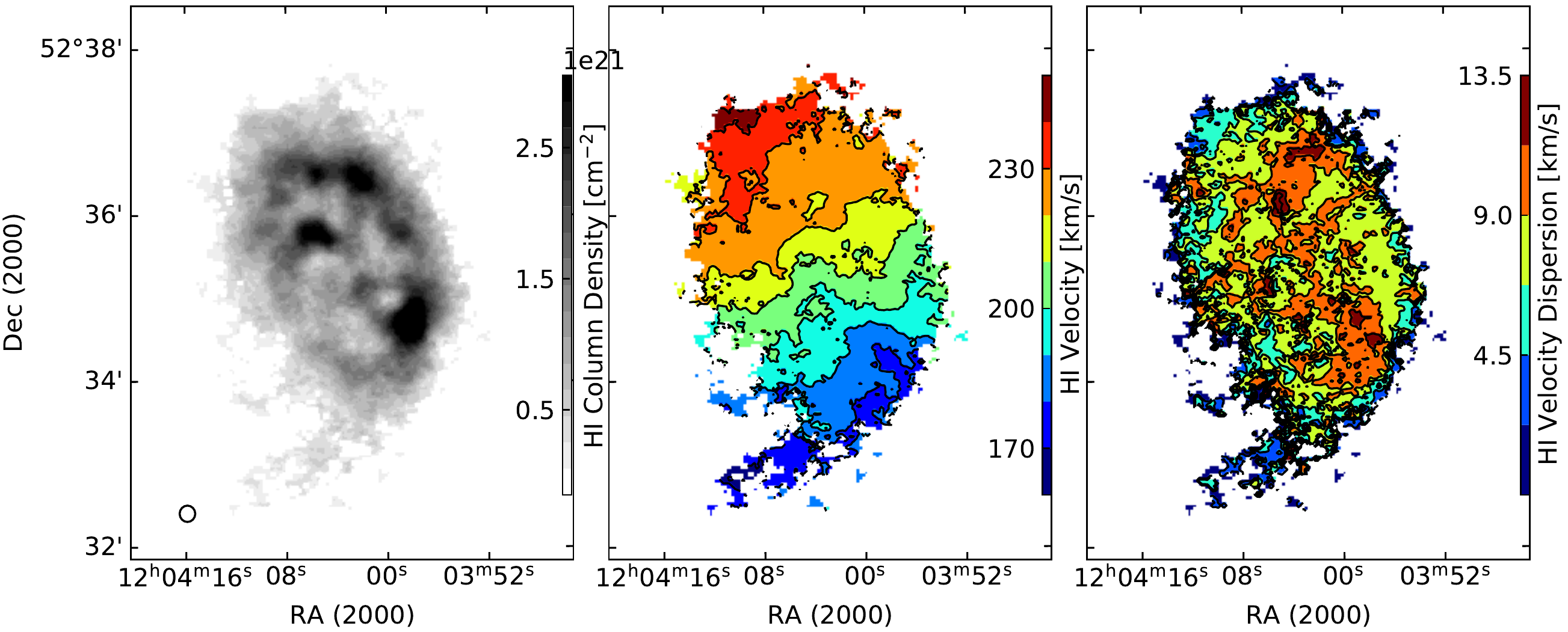}
        \caption{\small \textbf{NGC 4068}   HI moment maps from VLA observations Left: HI column density in 10$^{21}$ hydrogen atoms cm$^{-2}$, Center: HI velocity map with isovelocity contours spaces every 10 km s$^{-1}$, Right: HI velocity dispersion map with isovelocity contours at 2.25 km s$^{-1}$ spacing. The beam size (11.83''$\times$11.29'') of the   HI data cube used is shown in the bottom left of the left panel.}
    \label{4068_VLA}
\end{figure*}

\begin{figure*}[!th]
    \centering
    \includegraphics[width=.9\textwidth]{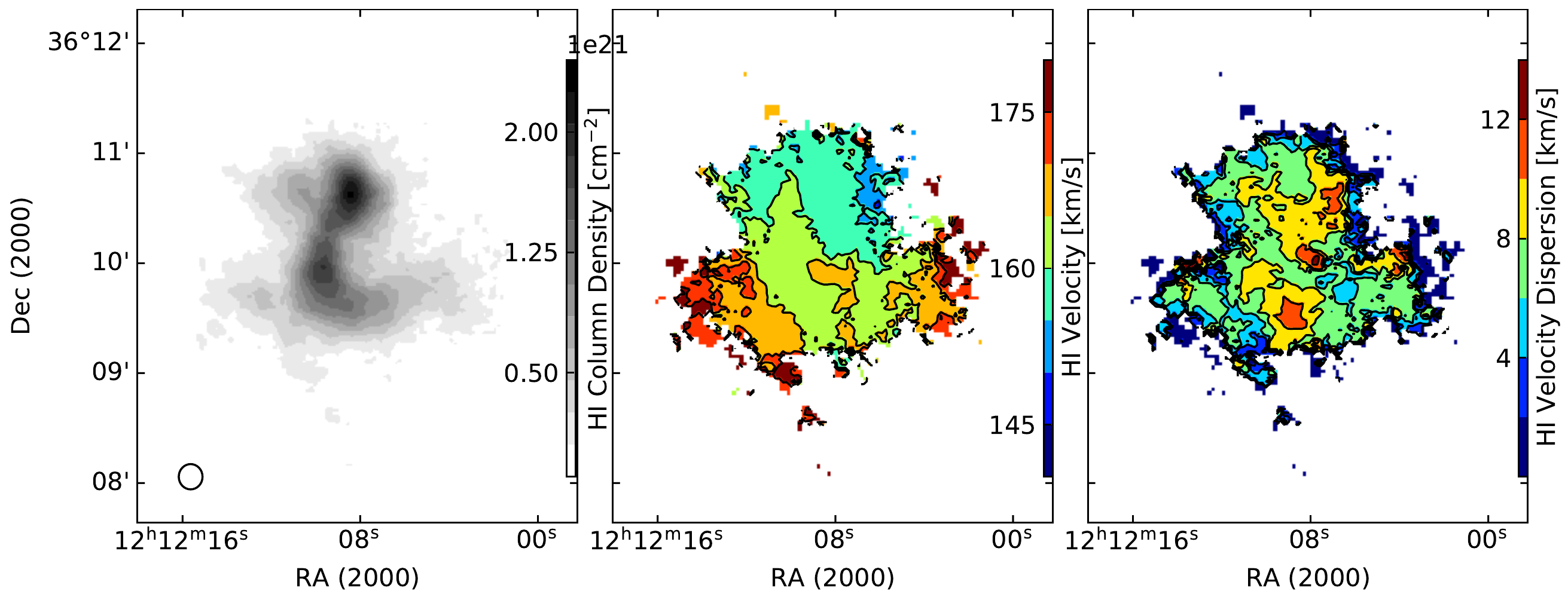}
        \caption{ \small \textbf{NGC 4163} HI moment maps from VLA observations Left: HI column density in 10$^{21}$ hydrogen atoms cm$^{-2}$, Center: HI velocity map with isovelocity contours spaces every 5 km s$^{-1}$, Right: HI velocity dispersion map with isovelocity contours at 2 km s$^{-1}$ spacing. The beam size (13.80''$\times$12.96'') of the   HI data cube used is shown in the bottom left of the left panel.}
    \label{4163_VLA}
\end{figure*}

\begin{figure*}[!th]
    \centering
    \includegraphics[width=.9\textwidth]{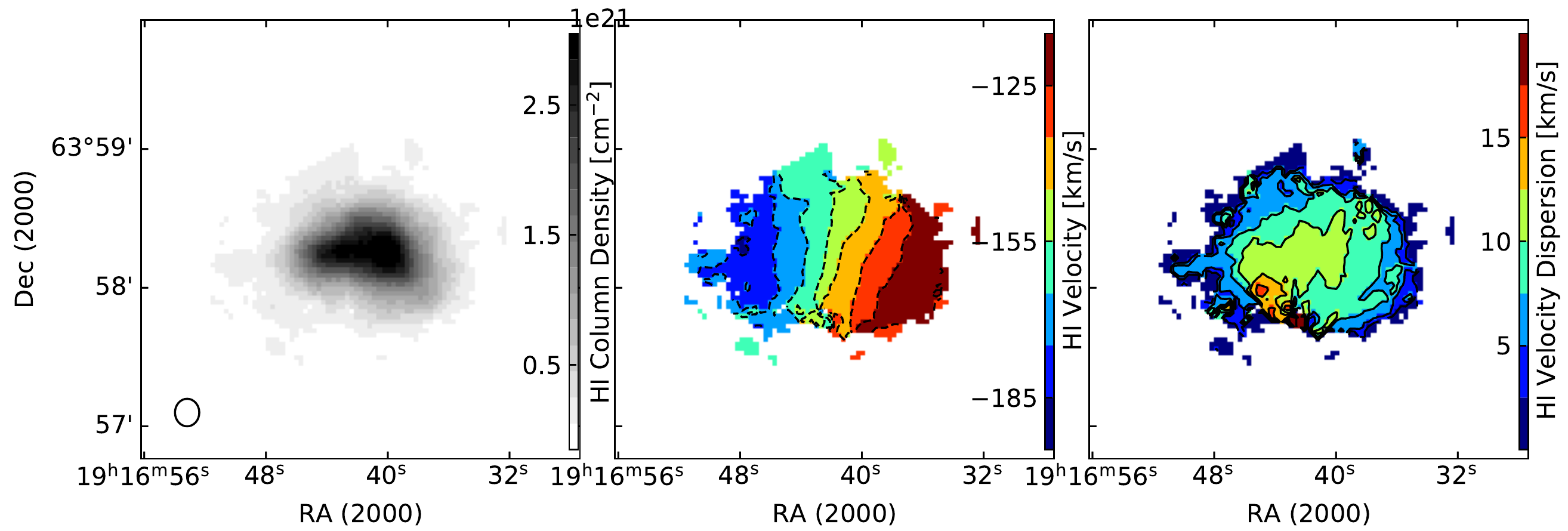}
        \caption{\small \textbf{NGC 6789}   HI moment maps from VLA observations Left:   HI column density in 10$^{21}$ hydrogen atoms cm$^{-2}$, Center: HI velocity map with isovelocity contours spaces every 10 km s$^{-1}$, Right: HI velocity dispersion map with isovelocity contours at 2.5 km s$^{-1}$ spacing. The beam size (11.92''$\times$10.52'') of the   HI data cube used is shown in the bottom left of the left panel.}
    \label{6789_VLA}
\end{figure*}

\begin{figure*}[!th]
    \centering
    \includegraphics[width=.9\textwidth]{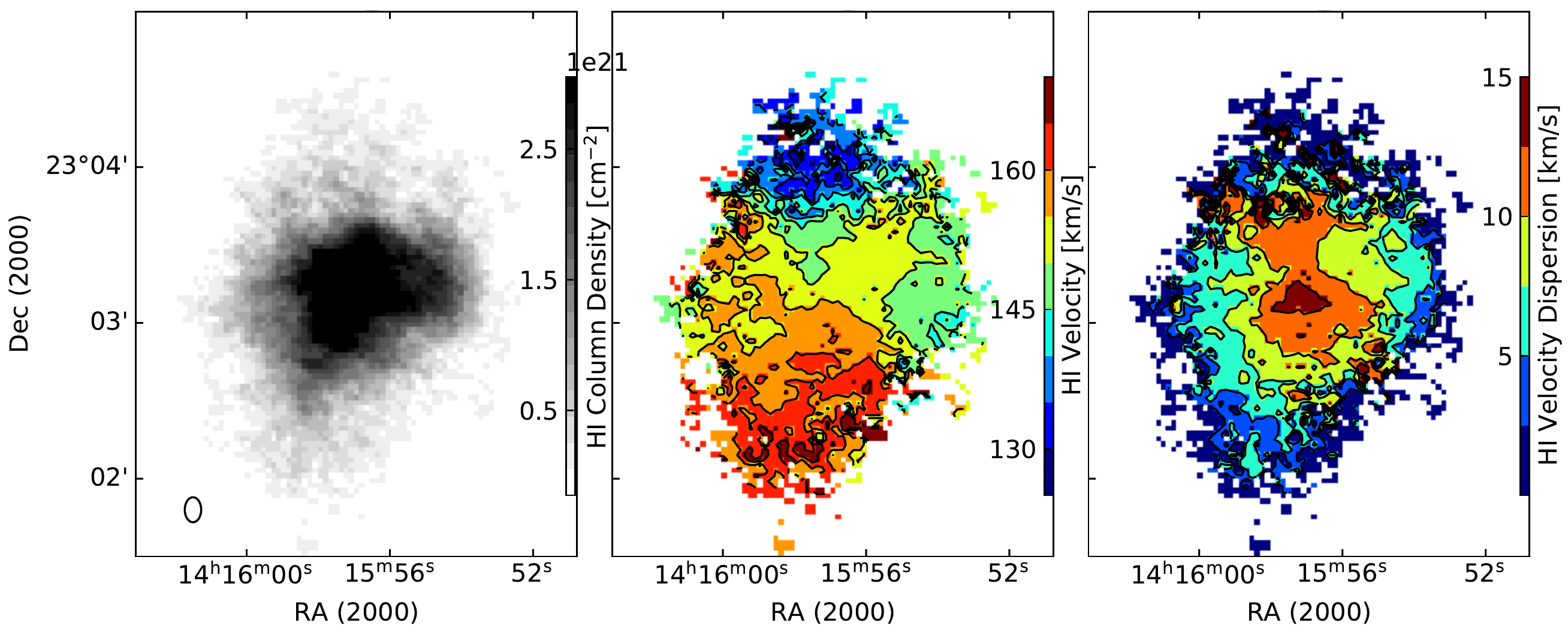}
        \caption{\small \textbf{UGC 9128} HI moment maps from VLA observations Left:   HI column density in 10$^{21}$ hydrogen atoms cm$^{-2}$, Center: HI velocity map with isovelocity contours spaces every 5 km s$^{-1}$, Right: HI velocity dispersion map with isovelocity contours at 2.5 km s$^{-1}$ spacing.  The beam size (9,64''$\times$6.35'') of the HI data cube used is shown in the bottom left of the left panel.}
    \label{9128_VLA}
\end{figure*}

For two of the galaxies (NGC 4163 and UGC 9128), archival VLA B, C, and D-configuration observations from VLA-ANGST \citep{VLAANGST}, and LITTLE THINGS \citep{LTHINGS}, were used.  For NGC 4068 and NGC 6789, we present new VLA observations -- B and C-configuration for NGC 6789 and B-Configuration for NGC 4068 data-- along with C-configuration data of NGC 4068 published in \cite{Richards18}(see Table \ref{table:hiobs} for new observations). For both galaxies the standard flux calibrator 3C286 was observed at the beginning of each observing block and the phase calibrator (either J1219+4829 or J2022+6136) was observed every 35 minutes to flux and phase calibrate the data.

For this study, the archival data for NGC 4163 and UGC 9128 were reprocessed to match the handling of NGC 4068 and NGC 6789.  Each set of new and archival data were loaded into \emph{AIPS}\footnote{The Astronomical Image Processing System (AIPS) was developed by the NRAO.} to be processed. The inner 75 percent of each observation block were combined to create a 'channel zero` for the data set, which was then flagged uniformally for radio frequency interference before flux and phase calibration. The calibration solutions were applied to the line data before it was bandpass calibrated using the flux calibrator. After calibrations were applied, the line data were corrected for not Doppler tracking with CVEL before being continuum subtracted in the \textit{uv} plane.  After Doppler correction and continuum subtraction, the individual observing blocks for each galaxy were combined and data cubes of multiple different resolutions for each galaxy were created using IMAGR. For NGC 6789 and NGC 4068, the channels were binned by 3 for a velocity resolution of $\sim$2.5 km s$^{-1}$. The natural weighted (robust of 5) data cubes were selected in this paper for analysis as we preferenced sensitivity over spatial resolution.  The natural weighted beam sizes resulted in multiple resolution elements per region of interest for each galaxy.  The parameters of the resulting data cube for each galaxy are presented in Table \ref{table:cubes}. The total HI column density, velocity field, and velocity dispersion maps for each galaxy are presented in Figures \ref{4068_VLA}, \ref{4163_VLA}, \ref{6789_VLA}, and \ref{9128_VLA}.  The velocity dispersion maps will be discussed in more detail in Section 3.3.

\begin{table*}[th!] 
    \centering
    \caption{HI Observations}
        \begin{tabular}{l c c c c c} 
        \hline
        \hline 
        Galaxy & Array  & Project & Dates & TOS (hrs) & Ch Sep (km s$^{-1}$) \\ [0.5ex] 
        \hline 
        NGC 4068 & B & 16A-172 & 2016 July 30, Aug 6, 19, 21, 22, 31  & 13.5 & 0.825  \\
        NGC 4068 & C & 16A-013 & 2016 April 22, 23 & 6.8 & 0.825 \\
        NGC 6789 & C & 16A-172 & 2016 April 10 & 1.62 & 0.825 \\
        NGC 6789 & B & 16A-172 & 2016 June 20, 30, July 5, 9 & 13.15 & 0.825  \\[1ex]
        \hline 
    \end{tabular} 
    \tablecomments{\footnotesize {C-Configuration data for NGC 4068 previously published in \cite{Richards18}}}
    \label{table:hiobs} 
\end{table*}

\begin{table*}[th!] 
    \centering
    \caption{HI Data Cubes and Properties}
        \begin{tabular}{c c c c c c c c c} 
        \hline
        \hline 
        Galaxy & $\Delta$v &  Beam & P.A. & RMS & HI flux & HI Mass \\
         & km s$^{-1}$ & arcsec$\times$arcsec & deg & mJy bm$^{-1}$ & Jy km s$^{-1}$  & log(M$_\odot$) \\ [0.5ex] 
        \hline 
        NGC 4068 & 2.47 & 11.83$\times$11.29 & 14.2 & 0.7095 & 40.1$\pm$4.0 & 8.30$\pm$0.04 \\
        NGC 4163 & 1.29 & 13.80$\times$12.96 & 24.9 & 0.839 & 9.9$\pm$.99 & 7.28$\pm$0.05 & \\
        NGC 6789 & 2.47 & 11.92$\times$10.52 & -64.8 & 0.536 & 4.9$\pm$.5 & 7.17$\pm$0.05 &\\
        UGC 9128 & 1.29 & 9.64$\times$6.35 & 79.4 & 0.865 & 15.3$\pm$1.5 & 7.25$\pm$0.05 &\\[1ex] 
        \hline 
    \end{tabular} 
    \label{table:cubes} 
\end{table*}

\subsection{Archival \textit{HST} Observations}

CMDs of resolved stellar populations from \textit{HST} observations taken with the Advanced Camera for Survey instrument (ACS: \citealt{ACS}) and the Wide Field Planetary Camera 2 instrument (WFPC2: \citealt{WFPC2}) were used to determine the SFHs.  Details of the observations are listed in Table  \ref{table:HST_obs}.  The ACS observations were taken of NGC 4068, NGC 4163, and UGC 9128 with the camera's F606W V filter and F814W I filter. The WFPC2 observations were taken of NGC 6789 with the F555W V filter and F814 I filter.  The ACS instrument has a 202"$\times$202" field of view with a native pixel scale of 0.05" pixel$^{-1}$ and the WFPC2 instrument has three 800$\times$800 pixel wide field CCDs, with a 0.1" pixel $^{-1}$ pixel scale, and a 800$\times$800 pixel planetary camera CCD with a 0.05" pixel $^{-1}$ pixel scale. 

The optical imaging was processed in an identical manner to that used in STARBIRDS \citep{McQuinn15a}. We provide a summary of the data reduction here and refer the reader to \cite{McQuinn10a} for a detailed description. Photometry was performed on the pipeline processed, charge transfer efficiency corrected images using the software HSTphot optimized for the ACS and WFPC2 instruments \citep{Dolphin00}. The photometry was filtered to include well-recovered point sources with the same quality cuts on signal-to-noise-ratios, crowding conditions, and sharpness parameters as applied in STARBIRDS. Artificial star tests were run on the individual images to measure the completeness of the stellar catalogs. As the derivation of the SFHs require a well-measured completeness function, we ran $\sim4M$ artificial star tests over each full field of view, ensuring sufficient number of stars in the individual smaller regions used in the analysis.

\begin{table*}[ht!] 
    \centering
    \caption{\textit{HST} Observations}
        \begin{tabular}{c c c c c c c} 
        \hline
        \hline 
        Galaxy & \textit{HST}  &  PI & Instrument & F555W & F606W & F814W \\
         & Proposal ID &  &  & sec & sec & sec \\ [0.5ex] 
        \hline 
        NGC 4068 & 9771 & Karachentsev & ACS & -- & 1200 & 900 \\
        NGC 4163 & 9771 & Karachentsev & ACS & -- & 1200 & 900 \\
        NGC 6789 & 8122 & Schulte-Ladbeck & WFPC2 & 8200 & -- & 8200 \\
        UGC 9128 & 10210 & Tully & ACS & -- & 990 & 1170 \\[1ex] 
        \hline 
    \end{tabular} 
    \label{table:HST_obs} 
\end{table*}

\begin{table}[h!] 
    \centering
    \caption{SparsePak Observations}
        \begin{tabular}{l c c c c c c} 
        \hline
        \hline 
        Galaxy & No. of & Date &  $\lambda$ Center  & ToS   \\
         & Fields & of Obs &  {\AA} & sec \\ [0.5ex] 
        \hline 
        NGC 4068 & 2 & April 3 2016 &  6680.530 & 1800 \\
        NGC 4163 & 1 & April 23 2017 & 6681.184 & 2700 \\
        NGC 6789 & 1 & April 2 2016 & 6680.530 & 1800 \\
        UGC 9128 & 1 & April 22 2017 & 6681.184 & 2700 \\[1ex] 
        \hline 
    \end{tabular} 
    \label{table:optobs} 
\end{table}
\vspace{-3mm}

\subsection{SparsePak Observations}

Spatially resolved spectroscopy of the ionized gas were taken with the SparsePak IFU \citep{Sparspak} on WIYN 3.5m telescope in April 2016 and April 2017. The SparsePak IFU has 82 4.69" diameter fibers arranged in a fixed 70"$\times$70" square, with the fibers adjacent to each other in the core and separated by 11" in the rest of the field. All observations were taken with the same bench set up  with the 316@63.4 bench spectrograph including the X19 blocking filter and an order 8 grating.  The resulting wavelength range was from 6480 {\AA} to 6890 {\AA} with a velocity resolution of 13.9 km s$^{-1}$ pixel$^{-1}$.  To fill in the gaps between fibers, a three pointing dither pattern was used. For each dither pointing three exposures of either 600 seconds (NGC 4068 and NGC 6789) or 900 seconds (NGC 4163 and UGC 9128) were taken in order to detect diffuse ionized gas, not just star forming regions.  For NGC 4068, two pointings were used to cover the full extent of the galaxy's H$\alpha$ emission on the sky.  Observations of blank sky were also taken to remove telluric line contamination, as the galaxies were more extended than the SparsePak field-of-view.

The SparsePak data were processed using the standard tasks in the \emph{IRAF}\footnote{IRAF is distributed by NOAO, which is operated by the Association of Universities for Research in Astronomy, Inc., under cooperative agreement with the National Science Foundation} HYDRA package.  The data were bias-subtracted, dark corrected, and cosmic ray cleaned, before the task DOHYDRA was used to fit and extract the apertures from the IFU data.  The spectra were wavelength calibrated using a solution created from Th-Ar lamp observations. The individual images were sky subtracted using a separate sky pointing, scaled to the 6577\AA \ telluric line. After sky-subtraction, the 3 exposures were averaged together to increase the signal-to-noise ratio.  A flux calibration using observations of spectrophotometric standards from \cite{Oke90} was applied to enable measurement of relative line strengths, although the nights were not photometric.  

\begin{figure*}[!th]
    \centering
    \includegraphics[width=.9\textwidth]{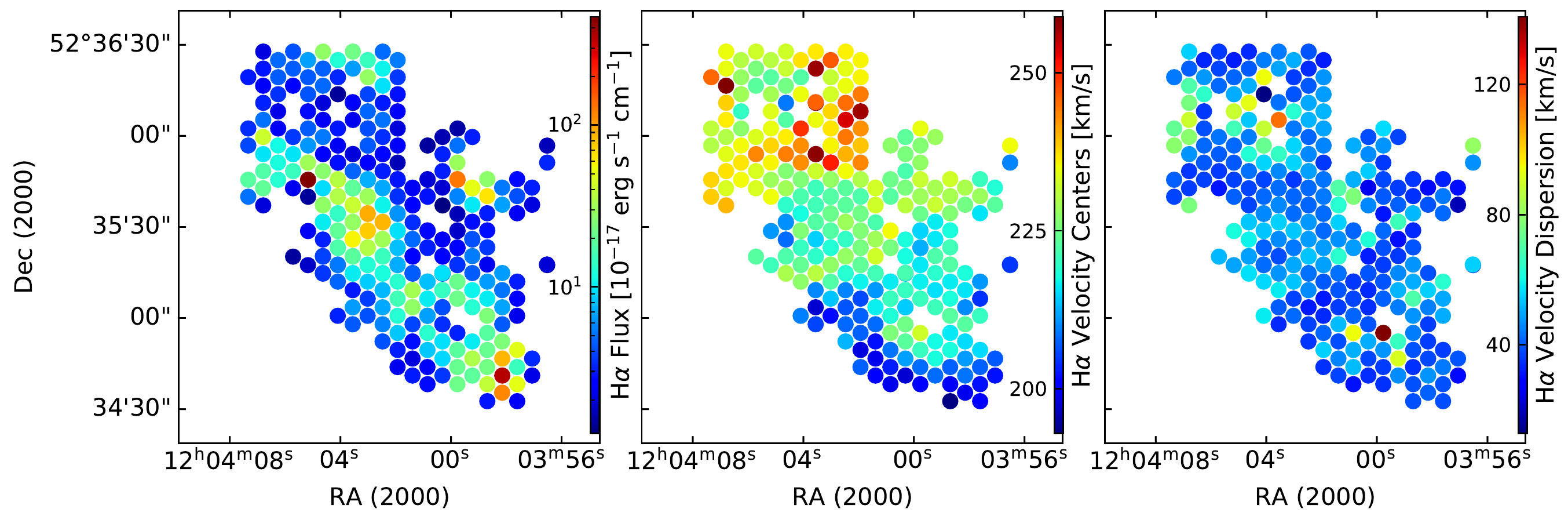}
        \caption{\small \textbf{NGC 4068} maps from observations with the SparsePak IFU on the WIYN 3.5m telescope, with H$\alpha$ line measurements from PAN.  Left: H$\alpha$ on a log scale in units of 10$^{-17}$ erg s$^{-1}$ cm$^{-1}$, Center: H$\alpha$ line centers map, Right: H$\alpha$ velocity dispersion ($\sigma_{H\alpha}$) map. Each filled circle corresponds to a fiber's size and position on the sky. For reference to where the Sparsepak fibers fall on the galaxy compared with the HI and optical distributions see Figure \ref{40683_frame}.}
    \label{4068_Sparse}
\end{figure*}

\begin{figure*}[!th]
    \centering
    \includegraphics[width=.9\textwidth]{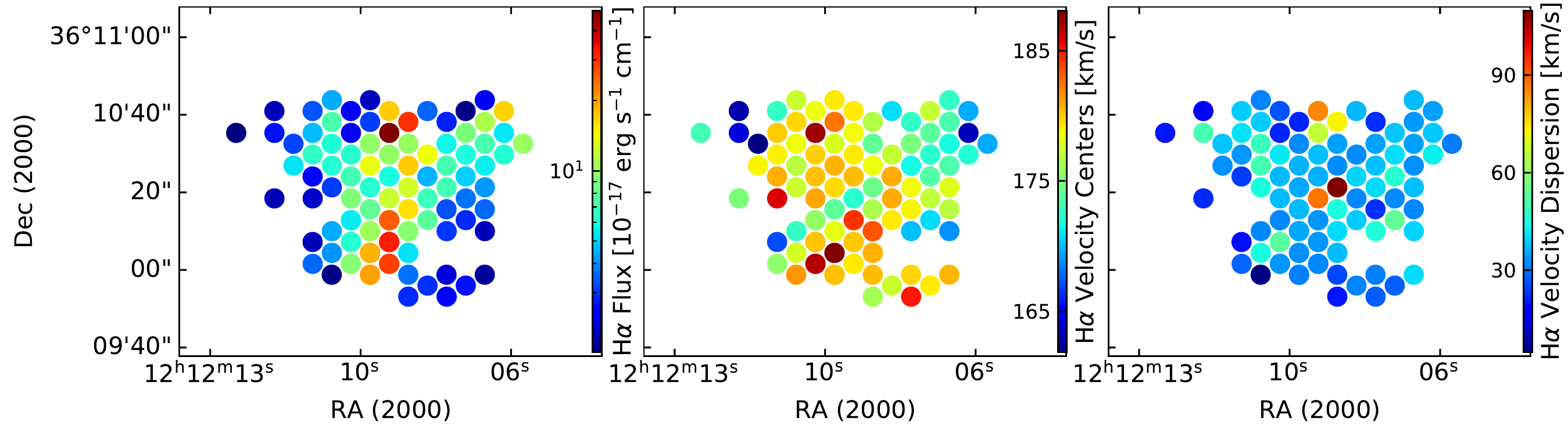}
        \caption{ \small \textbf{NGC 4163} maps from observations with the SparsePak IFU on the WIYN 3.5m telescope, with H$\alpha$ line measurements from PAN.  Left: H$\alpha$ on a log scale in units of 10$^{-17}$ erg s$^{-1}$ cm$^{-1}$, Center: H$\alpha$ line centers map, Right: H$\alpha$ velocity dispersion ($\sigma_{H\alpha}$) map. Each filled circle corresponds to a fiber's size and position on the sky. For reference to where the Sparsepak fibers fall on the galaxy compared with the HI and optical distributions see Figure \ref{41633_frame}.}
    \label{4163_Sparse}
\end{figure*}

\begin{figure*}[!th]
    \centering
    \includegraphics[width=.9\textwidth]{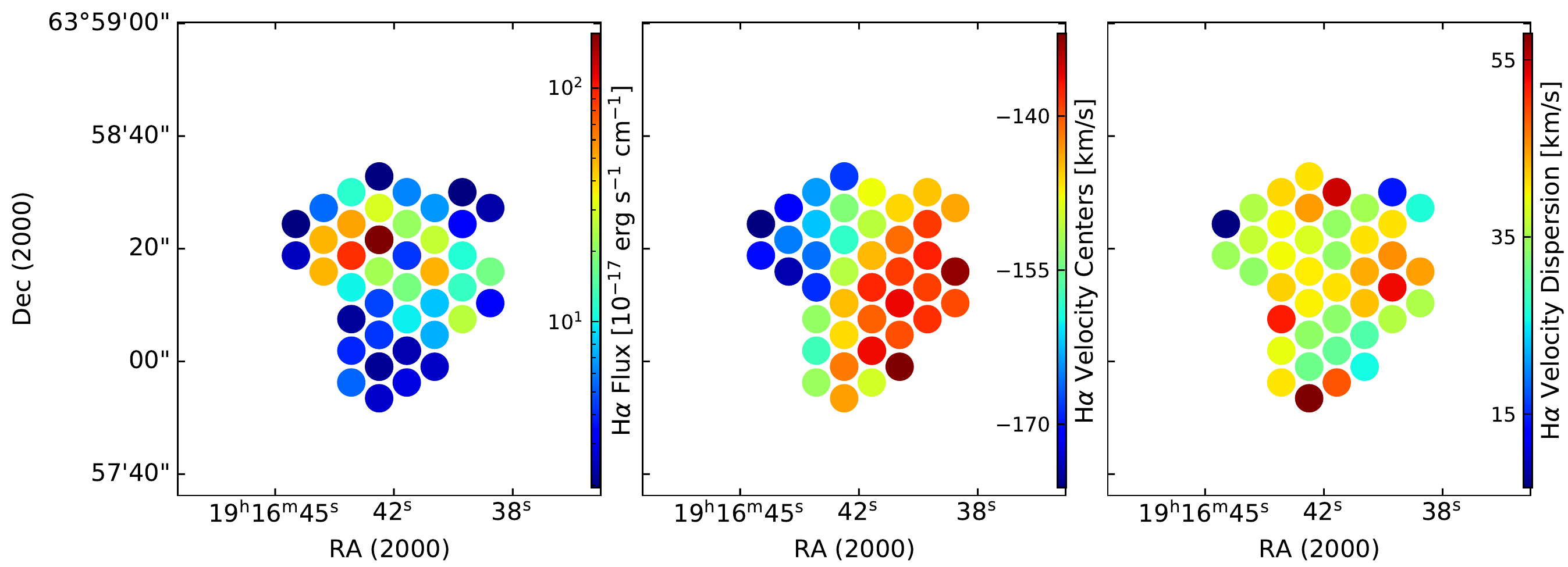}
        \caption{\small \textbf{NGC 6789} maps from observations with the SparsePak IFU on the WIYN 3.5m telescope, with H$\alpha$ line measurements from PAN.  Left: H$\alpha$ on a log scale in units of 10$^{-17}$ erg s$^{-1}$ cm$^{-1}$, Center: H$\alpha$ line centers map, Right: H$\alpha$ velocity dispersion ($\sigma_{H\alpha}$) map.  Each filled circle corresponds to a fiber's size and position on the sky. For reference to where the Sparsepak fibers fall on the galaxy compared with the HI and optical distributions see Figure \ref{67893_frame}.}
    \label{6789_Sparse}
\end{figure*}

\begin{figure*}[!th]
    \centering
    \includegraphics[width=.9\textwidth]{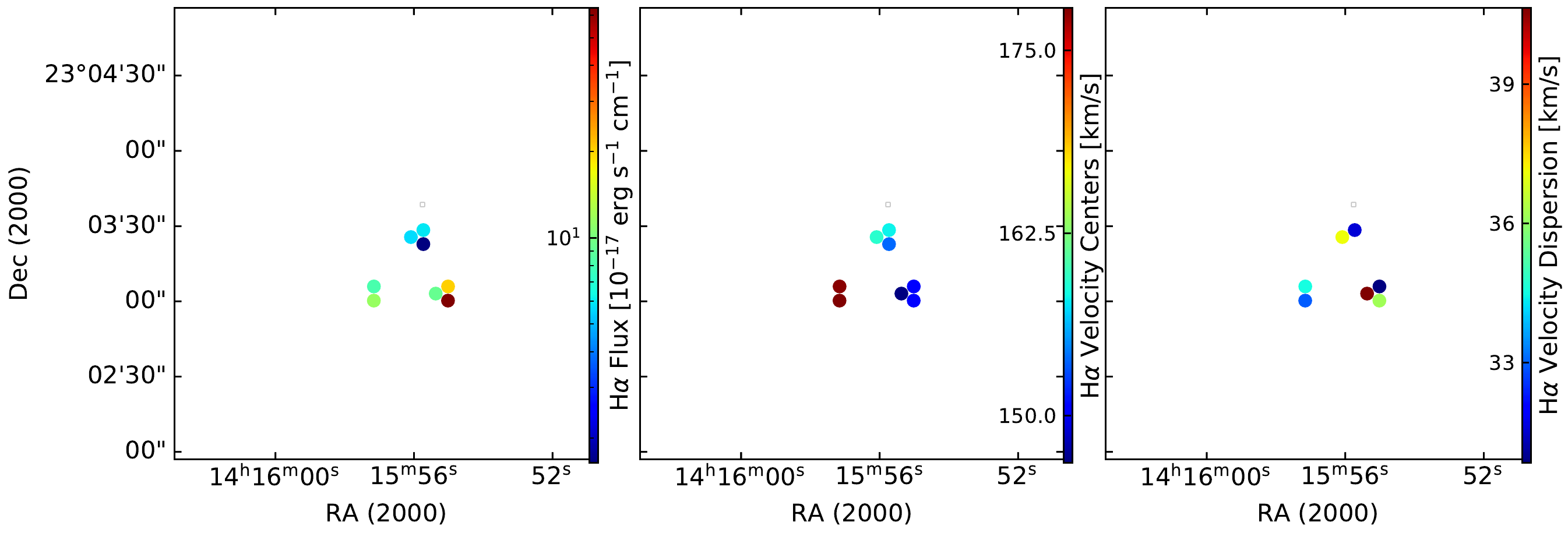}
        \caption{\small \textbf{UGC 9128} maps from observations with the SparsePak IFU on the WIYN 3.5m telescope, with H$\alpha$ line measurements from PAN.  Left)H$\alpha$ on a log scale in units of 10$^{-17}$ erg s$^{-1}$ cm$^{-1}$, Center) H$\alpha$ line centers map, Right) H$\alpha$ velocity dispersion ($\sigma_{H\alpha}$) map. Each filled circle corresponds to a fiber's size and position on the sky. For reference to where the Sparsepak fibers fall on the galaxy compared with the HI and optical distributions see Figure \ref{91283_frame}}
    \label{9128_Sparse}
\end{figure*}

The galaxy spectra were smoothed by 1 pixel (0.306\AA) in order to improve S/N. For each fiber spectra, the emission lines were fit to Gaussians using the IDL software suite Peak ANalaysis (PAN; \citealt{PAN}). The measured H$\alpha$ line widths were corrected for instrumental broadening of 47.8$\pm$1.6 km s$^{-1}$, as measured from the equivalently smoothed ThAr spectra.  The H$\alpha$ line fluxes, centers, and velocity dispersions from PAN where visually inspected and the fiber positions that passed were placed into a grid mapping their SparsePak fiber placements. The output line fluxes, centers, and velocity dispersions are shown in Figures \ref{4068_Sparse}, \ref{4163_Sparse}, \ref{6789_Sparse}, and \ref{9128_Sparse}.

\section{Region Processing} \label{regions}

In order to study the spatially resolved impact of star formation on the ISM, the galaxies were divided into regions of interest with a set physical size. For each region the SFH, ionized gas velocity dispersion, and atomic gas velocity dispersions are measured independently.

\subsection{Galaxy Divisions}

Two competing criteria were balanced to determine an appropriate physical scale for the analysis: the number of star counts within each region and retaining information about the local effects of stellar feedback on the ISM.  In other words, regions must be large enough to ensure reliable SFHs with sufficient time resolution ($\simeq$ 25 Myrs in the most recent time bins) and a 500 Myrs baseline, while being small enough that any local turbulence effects are not washed out. We chose to partition each galaxy into square regions with a physical size of 400 pc per side as a compromise between observational limits and theoretical expectations. 

A region size of 400$\times$400 pc was determined as the largest reasonable scale for the analysis from work on clustered SNe.  As individual and clustered SNe (superbubbles) are likely the most important mechanism for driving turbulence in the ISM \citep{Norman96,OstrikerShetty11,Kimetal11,ElBrady19}, the maximum physical scale was limited to the scale of these events.  From simulations and theory, the predicted range over which superbubbles input momentum into the ISM is about one to a few times the scale height of the galaxy \citep{Kimetal17,Gentry17}, or roughly 200 to 600 pc for dwarf galaxy disk thicknesses \citep{Bacchini20b}.  Thus, region sizes larger than 400 pc would not be able to identify the impacts of the local star formation activity from the global star formation activity on the ISM. 

\begin{table}[th!] 
    \centering
    \caption{Galaxy Region Sizes}
        \begin{tabular}{c c c} 
        \hline
        \hline 
        Galaxy & Region & Region \\
          & arcsec$\times$arcsec & pc$\times$pc \\ [0.5ex] 
        \hline 
        NGC 4068 & 18$\times$18 & 405$\times$405\\
        NGC 4163 & 28$\times$28 & 390$\times$390\\
        NGC 6789 & 24$\times$24 & 413$\times$413\\
        UGC 9128 & 38$\times$38 & 407$\times$407\\[1ex] 
        \hline 
    \end{tabular} 
    \label{table:regions} 
\end{table}

\begin{figure*}[!th]
    \centering
    \includegraphics[width=.9\textwidth]{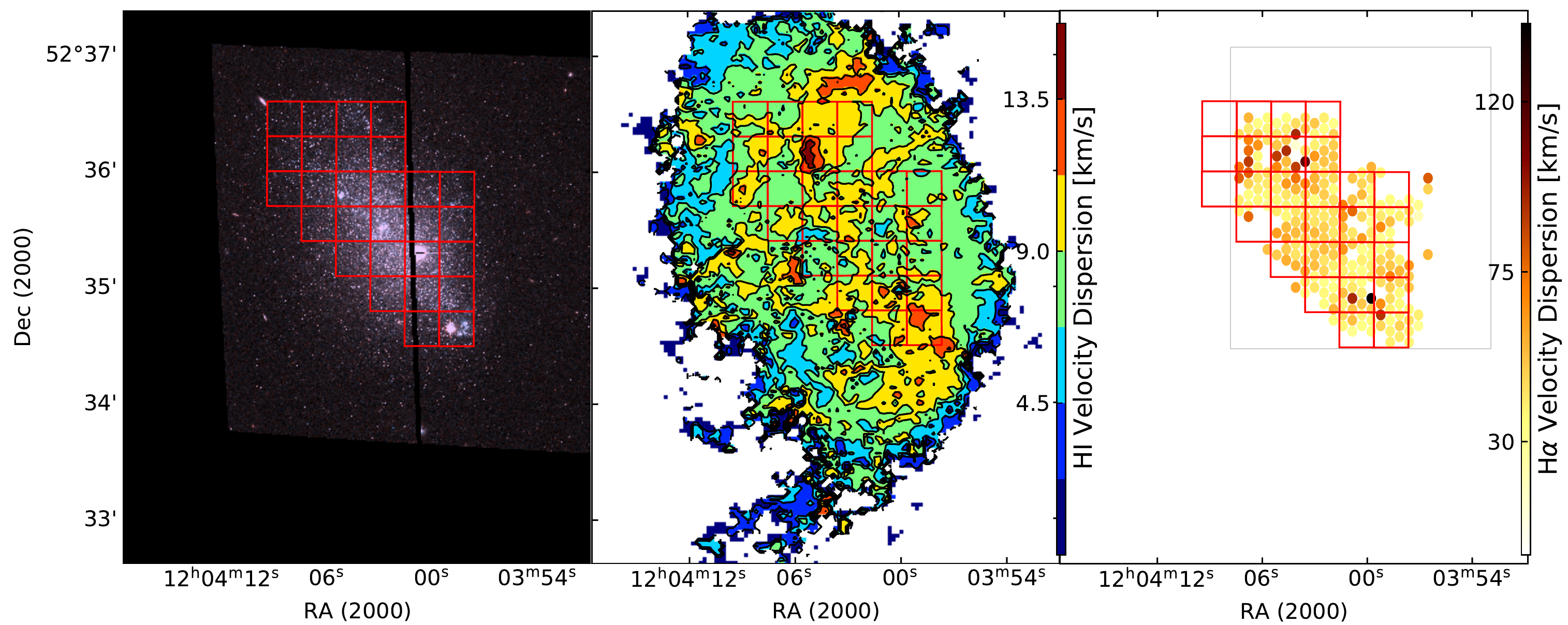}
        \caption{\small \textbf{NGC 4068} Left) Two color image from \textit{HST} F814W (red) and F606W (blue) observations with ACS, Center) HI dispersion map from VLA observations with isovelocity contours in 2.25 km s$^{-1}$ step size, Right) $\sigma_{H\alpha}$ map from the SparsePak IFU on the WIYN 3.5m telescope, with each filled circle corresponding to a fiber's size and position on the sky. Overlaid on all three panels are the outlines of the regions used for the analysis.}
    \label{40683_frame}
\end{figure*}

\begin{figure*}[!th]
    \centering
    \includegraphics[width=.9\textwidth]{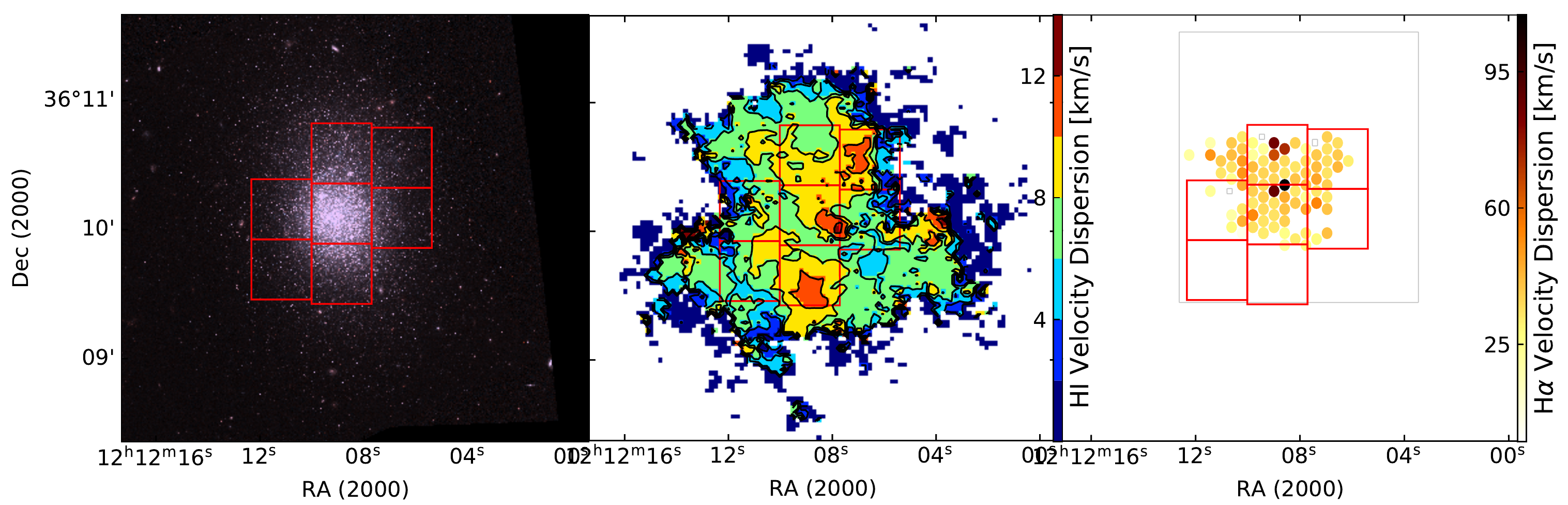}
        \caption{ \small \textbf{NGC 4163} Left) Two color image from \textit{HST} F814W (red) and F606W (blue) observations with ACS, Center) HI dispersion map from VLA observations with isovelocity contours in 2 km s$^{-1}$ step size, Right) $\sigma_{H\alpha}$ map from the SparsePak IFU on the WIYN 3.5m telescope, with each filled circle corresponding to a fiber's size and position on the sky. Overlaid on all three panels are the outlines of the regions used for the analysis.}
    \label{41633_frame}
\end{figure*}

\begin{figure*}[!th]
    \centering
    \includegraphics[width=.9\textwidth]{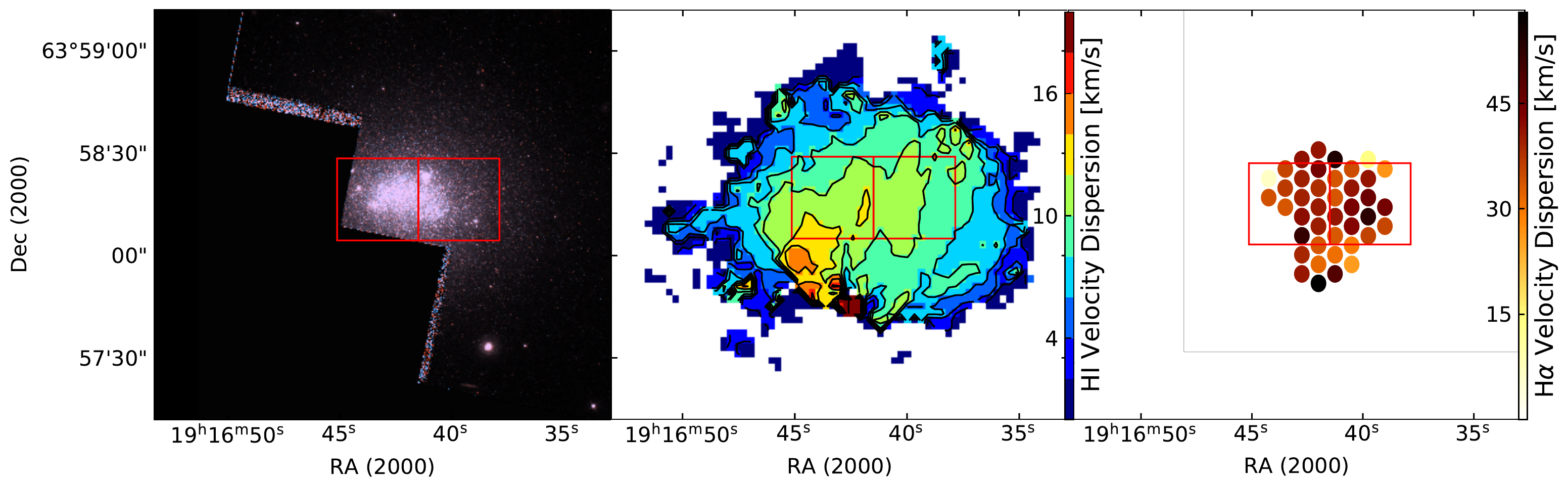}
        \caption{\small \textbf{NGC 6789} Left) Two color image from \textit{HST} F814W (red) and F555W (blue observations with WFPC2), Center) HI dispersion map from VLA observations with isovelocity contours in 2 km s$^{-1}$ step size, Right) $\sigma_{H\alpha}$ map from the SparsePak IFU on the WIYN 3.5m telescope, with each filled circle corresponding to a fiber's size and position on the sky. Overlaid on all three panels are the outlines of the regions used for the analysis.}
    \label{67893_frame}
\end{figure*}

\begin{figure*}[!th]
    \centering
    \includegraphics[width=.9\textwidth]{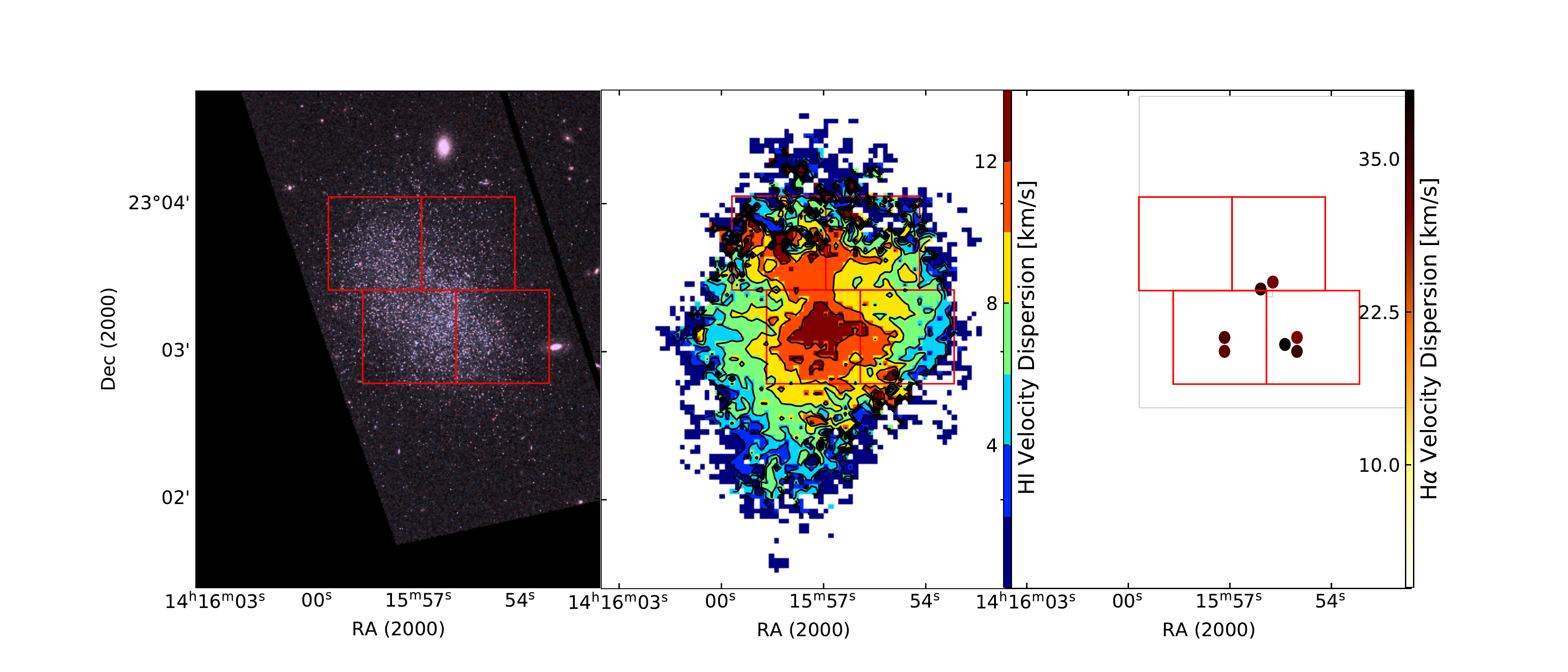}
        \caption{\small \textbf{UGC 9128} Left)  Two color image from \textit{HST} F814W (red) and F606W (blue) observations with ACS, Center) HI dispersion map from VLA observations with isovelocity contours in 2 km s$^{-1}$ step size, Right) $\sigma_{H\alpha}$ map from the SparsePak IFU on the WIYN 3.5m telescope, with each filled circle corresponding to a fiber's size and position on the sky. Overlaid on all three panels are the outlines of the regions used for the analysis.}
    \label{91283_frame}
\end{figure*}

For each galaxy, the angular size of the regions was calculated based off their distance and rounded to the nearest 2'', the pixel size of the HI data. The physical and angular size of each region is listed in Table \ref{table:regions}.  Regions were arranged as a grid across the galaxies with grid placement adjusted to maximize the number of regions with reliable SFHs over the past 500 Myrs, and to have as consistent a HI velocity dispersion within the region --based off HI second moment maps-- as possible. Adjustments to the region placements from a simple grid were made with the goal of measuring the SFH in regions with particularly high or low HI or H$_\alpha$ velocity dispersion along with adjustments based off the stellar distribution. The final region placements for each galaxy are shown in Figures \ref{40683_frame}, \ref{41633_frame}, \ref{67893_frame}, and \ref{91283_frame}.

\subsection{Star Formation Histories}\label{sec:SFH}

The numerical CMD fitting program, MATCH was utilized to reconstruct SFHs from resolved stellar populations \citep{Dolphin02}.  To summarize, MATCH uses an assumed initial mass function (IMF) along with a stellar evolution library to create a series of synthetic simple stellar populations (SSPs) with different ages and metallicities. A large number of synthetic CMDs were produced for each region with each CMD containing stars with limited ranges of age (0.05 dex) and metallicity (0.10 dex). The SFH solutions were based off a Kroupa IMF \citep{Kroupa01}, an assumed binary function of 35\% with a flat binary mass ratio distribution, and the PARSEC stellar library \citep{Bressan12}.  We assumed no internal differential extinction because, for the low-masses of this sample of galaxies, internal extinction should be low (i.e., the mass-metallicity relation; \citealt{Berg12}). Observational errors (from photon noise and blending) are simulated by using the completeness, photometric bias, and photometric scatter (all functions of color and magnitude) measured in artificial star tests. These synthetic CMDs, as well as simulated CMDs of foreground stars, were combined linearly to calculate the expected distribution of stars on the CMD for any SFH. 

\begin{figure*}[!th]
    \centering
    \includegraphics[width=.9\textwidth]{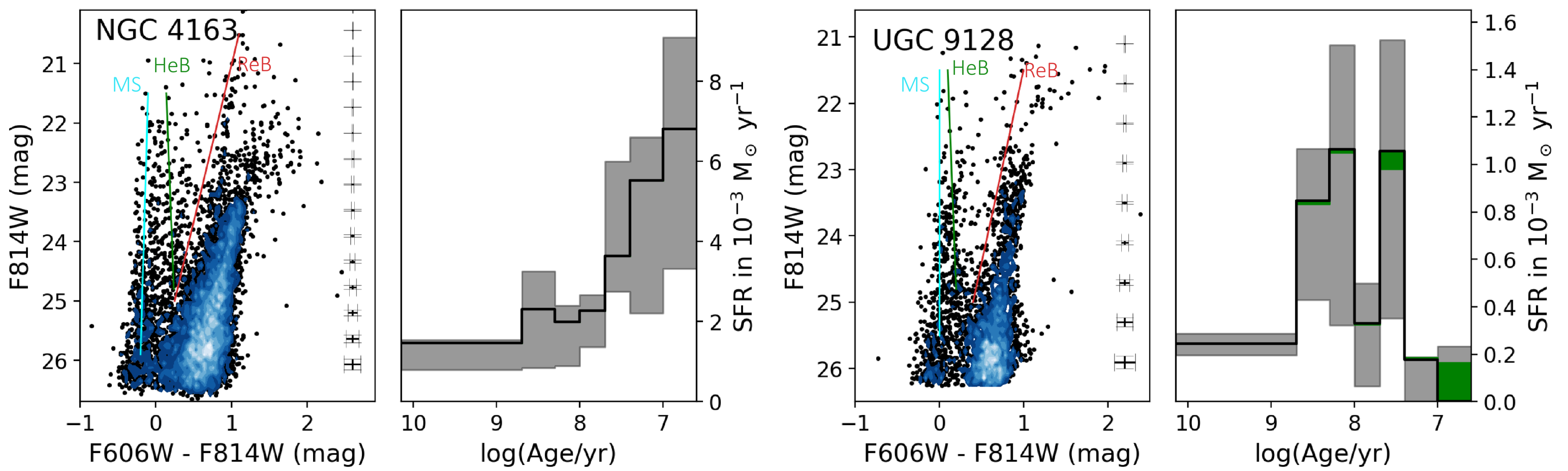}
        \caption{\small Example CMDs and SFHs for 400$\times$400 pc regions in UGC 9128 and NGC 4163 WFPC2 and ACS data. For both regions, the MS (blue), blue Helium burning stars (HeB) (green), and red HeB (red) sequences are traced.  The SFHs have $\leq$25 Myr time resolution in the two most recent time bins with $\Delta$log(t)=0.3 time steps and covers the 500 Myr baseline necessary for our science goals.  The grey shading is the combined systematic and random uncertainties on the SFR. The green shading represents the random uncertainties. In regions where no green shading is shown these uncertainties are too small to be seen on the scale of the figure. The CMD-derived SFHs are compared with regional measurements of the HI and H$\alpha$ turbulence to determine the time over which stellar feedback impacts multiple phases of the ISM.}
    \label{CMD}
\end{figure*}

With the synthetic and observed V vs (V-I) CMDs, the likelihood that the observed data were produced by the SFH of a particular synthetic CMD was calculated. A maximum likelihood algorithm was used to determine the SFH most likely to have produced the observed data for each region.  Systematic uncertainties from the stellar evolution models were estimated by applying shifts in luminosity and temperature to the observed stellar populations through Monte Carlo simulations \citep{Dolphin12}. Random uncertainties were estimated by applying a hybrid Markov Chain Monte Carlo simulation \citep{Dolphin13}.  The resulting CMD-based SFH provide $\Delta$log(t)=0.3 time resolution with a 500 Myr baseline. Example CMDs and the resulting SFHs for 400$\times$400 pc regions in NGC 4163, and UGC 9128 are show in in Figure \ref{CMD}.  A more complete description of the methods applied can be found in \cite{McQuinn10a} and the references therein. 

Dynamical studies indicate that low-mass galaxies are solid body rotators (e.g., \citealt{Skillman88,Skillman96_aspc,VanZee01}, see also the velocity fields in Figures \ref{4068_VLA} through \ref{9128_VLA}), which results in less radial and azimuthal mixing of stellar populations compared to the differential rotation of larger galaxies, allowing the SFHs of specific locations to be recovered. However, stellar populations do diffuse slowly, destroying the substructure made by the clusters, groups, and associations that they were born in (\citealt{Bastian10}, and references therein).  For dwarf galaxies, substructures can persist on timescales of $\sim$80 Myrs for the SMC \citep{Gieles08} to  $>$ 300 Myrs for DDO 165 \citep{Bastian11}.  As these timescales were measured down to the limiting depth of the photometery of the images, the stellar structures can not be probed on longer timescales, making these estimates \textit{lower limits}. Based off these lower limits we anticipate that we are accurately recovering the SFH of each region back more than 250 Myrs, and likely back 500 Myrs, the limit of the SFHs derived from the CMDs here.

\subsection{Turbulence Measurements} \label{turb}

To determine the turbulence for each region, two independent measures of the velocity dispersion and energy surface density of the HI were used, along one method for measuring the velocity dispersion of the ionized gas.  For the HI, the velocity dispersion of the region was characterized using moment maps (See Section \ref{mmaps}), which provides an estimate of the HI kinematics and makes no assumption about the underlying HI emission profile (e.g, \citealt{Tamburro09}). However, second moment measurements can be strongly effected by small amounts of gas at atypical velocities.  Independent of the moment maps, superprofiles were made using methods similar to \cite{Ianjama12} and \cite{Stilp13a} by co-adding line-of-sight profiles after correcting for rotational velocities (See Section \ref{sprofile}).  For the H$\alpha$ line emission, we measured the intensity weighted average velocity dispersion within each region (See Section \ref{havdisp}).

\subsubsection{Moment Maps}\label{mmaps}

The HI synthesis data cubes were processed with standard tools from the \emph{GIPSY} software package \citep{GIPSY} to extract the intensity weighted velocity dispersion maps of the four galaxies. To create the second moment maps, individual channels of the data cubes were smoothed by a factor of 2, and clipped at the 2$\sigma$ level, before being interactively blotted to identify signal.   Figures \ref{40683_frame}, \ref{41633_frame}, \ref{67893_frame}, and \ref{91283_frame} show the final velocity dispersion maps with the regions placements overlaid.  The flux weighted average of the second moment map was measured for each region: 

\begin{equation}
    \sigma_{m2}=\frac{\Sigma_i \sigma_i N_{HI,i}}{\Sigma_i N_{HI,i}}
\end{equation}
where N$_{HI,i}$ is the HI column density, and $\sigma_i$ is the second moment velocity dispersion of each pixel.  The column density weighted average velocity dispersions from the second moment maps are shown in Table \ref{vdisps}. For the uncertainty of the second moment velocity dispersion, the standard deviation of a weighted average was used.

\subsubsection{Superprofiles}\label{sprofile}

Superprofiles of the total HI flux within each region were constructed using techniques similar to those described in \cite{Ianjama12} to determine the velocity dispersion. Before summing the HI profiles within each region, the bulk motion of the gas was accounted for by shifting the individual profiles to a reference velocity of zero. The location of the peak for the individual profiles was estimated using the \emph{GIPSY} task XGAUFIT to fit each profile with a third order (\textit{h3}) Gauss-Hermite polynomial. A Gauss-Hermite \textit{h3} polynomial gives a robust estimate of the peak location even in the presence of asymmetries as it fits the skewness of the line profile (see \citealt{deBlok08} for details).  Line profiles were excluded from fitting if the maximum was less than 3$\sigma$ above the mean rms noise level per channel, or the velocity dispersion was less than the channel width to avoid fitting noise peaks.  After determining the center with XGAUFIT, SHUFFLE was used to shift the profiles to a reference velocity of zero.  The total flux within each region was calculated using the task FLUX after the lines were shifted. The uncertainty of each point in the super profiles is defined as:

\begin{equation}
    \sigma = \sigma_{ch,rms} \times \sqrt{N_{pix}/N_{pix/beam}}
    \label{uncereq}
\end{equation}
where $\sigma_{ch,rms}$ is the mean rms noise level per channel, N$_{pix}$ is the number of pixels contributing to a given point in the superprofile, and N$_{pix/beam}$ is the number of profiles in one resolution elements or pixels per beam size.

\begin{figure}
    \centering
    \includegraphics[width=0.45\textwidth]{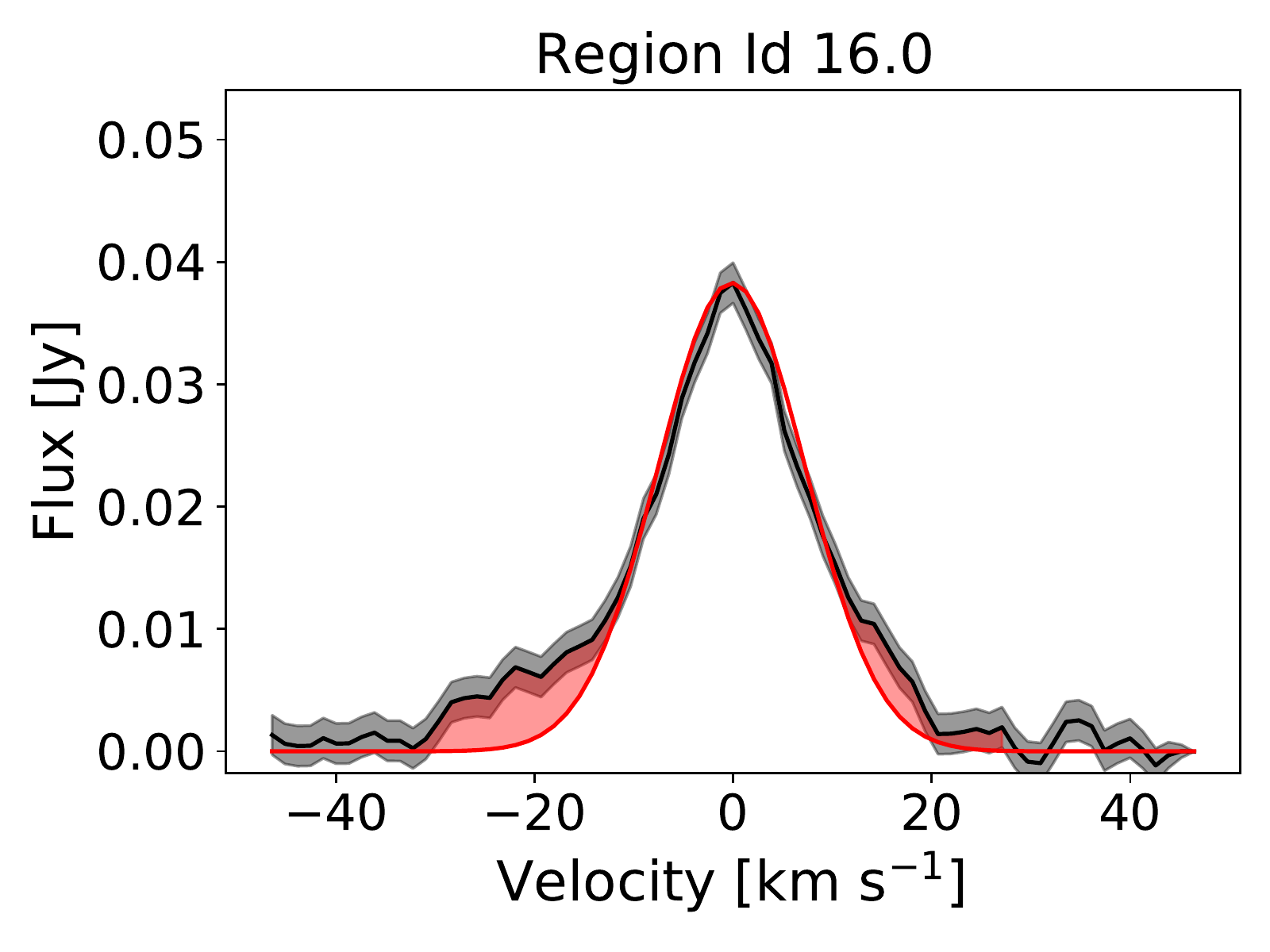}
    \caption{The superprofile of a selected region in NGC 4163.  The black line is the bulk-motion corrected HI flux from the region and the red line is the Gaussian fit for the data.  The shaded grey region is the error on the data, while the shaded red region is the wings of the HI flux.  The wings are the high velocity, low density gas that is poorly fit by a single Gaussian.}
    \label{super_pro}
\end{figure}

As a single Gaussian does not fit the low density HI flux at higher velocities well, we did not perform a traditional $\chi^2$ minimization to fit the line profiles. Instead, the process described in \cite{Stilp13a,Stilp13b} was used.  For each superprofile a Gaussian was scaled to the amplitude and the full-width at half-maximum (FWHM) of the line profile.  The HI flux at higher velocities and lower densities that is above the Gaussian fit is described as the wings of the superprofile (Figure \ref{super_pro}).  From the scaled Gaussian fits three parameters were measured:

\begin{enumerate}
    \item $\sigma_{central}$: the width of the scaled-Gaussian profile fit to the FWHM and amplitude of the observed HI superprofile.  $\sigma_{central}$ was chosen instead of FWHM as other studies often describe line width in term of a Gaussian $\sigma$ (e.g. \citealt{Ianjama12}).  
    \item $f_{wings}$: the fraction of HI in the wings of the profile where f$_{wings}$ is a measure of the fraction of gas moving at faster velocities than is expected compared to the bulk of the HI.
    \begin{equation}
        f_{wings}=\frac{\Sigma_{|v|> v_h}[S(v)-G(v)]}{\Sigma_{|v|>0}S(v)}
    \end{equation}
    where v is the center velocity of the profile, $v_h$ is the velocity at the half-width at half-maximum and $|v|> v_h$ is where the absolute value of the velocity is greater than the velocity at the half-width at half-maximum , S(\textit{v}) is the superprofile of the observed HI flux within the regions, and G(\textit{v}) is the scaled-Gaussian profile fit to the observed HI superprofile. 
    \item $\sigma^2_{wings}$:  the rms velocity of the HI flux in the profile wings, weighted by the fraction of gas in the observed HI superprofile S(\textit{v}) moving faster than the scaled-Gaussian  profile G(\textit{v}) predicts, used to characterize the velocity of the excess low density gas
    \begin{equation}
        \sigma^2_{wings}=\frac{\Sigma_{|v|> v_h}[S(v)-G(v)]v^2}{\Sigma_{|v|> v_h}[S(v)-G(v)]}
    \end{equation}
\end{enumerate}
To estimate the errors on these parameters, we assumed the observed superprofile is correct, and added Gaussian noise to each point based off Equation \ref{uncereq}.  The ``noisy'' data was refit with a Gaussian performing a standard $\chi^2$ minimization.  The process was repeated the 3000 times.  We took the 1$\sigma$ standard deviation of the refitting as the uncertainty on the superprofile parameters. The $\sigma_{cen}$ and $\sigma_{wing}$, along with $\sigma_{mom2}$, are listed below in Table \ref{vdisps} with their errors and region ID number and galaxy. 

\startlongtable
\begin{deluxetable*}{c c c c c c c c} 
    \tablecaption{Galaxy Region Velocity Dispersions}
    \tablehead{\colhead{Galaxy} & \colhead{Region ID} & \colhead{HI Surface Density} & \colhead{$\sigma_{mom2}$} & \colhead{$\sigma_{cen}$} & \colhead{$\sigma_{wings}$} & \colhead{f$_{wings}$} & \colhead{$\sigma_{H\alpha}$ $^A$} \\ 
    \colhead{} & \colhead{} & \colhead{(M$_\odot$ pc$^{-2}$)} & \colhead{(km $s^{-1}$)} & \colhead{(km $s^{-1}$)} & \colhead{(km $s^{-1}$)} & \colhead{} & \colhead{(km $s^{-1}$)} } 
    \startdata
        NGC 4068 & 14 & 13.6$\pm$1.4 & 9.4$\pm$1.2 & 11.0$\pm$0.3 & 29.0$\pm$1.6 & 0.065$\pm$0.014 & - \\
        & 15 & 14.8$\pm$1.5 & 8.9$\pm$1.4 & 9.9$\pm$0.2 & 26.4$\pm$1.2 & 0.070$\pm$0.012 & - \\
        & 16 & 15.8$\pm$1.6 & 8.7$\pm$0.6 & 9.5$\pm$0.2 & 24.3$\pm$0.8 & 0.065$\pm$0.010 & - \\
        & 24 & 17.0$\pm$1.7 & 9.1$\pm$1.0 & 9.7$\pm$0.2 & 31.5$\pm$0.7 & 0.142$\pm$0.010 & 48$\pm$24 \\
        & 25 & 21.0$\pm$2.1 & 8.3$\pm$0.7 & 8.6$\pm$0.1 & 29.1$\pm$1.3 & 0.073$\pm$0.009 & 40$\pm$16 \\
        & 26 & 11.9$\pm$1.2 & 9.0$\pm$1.3 & 10.1$\pm$1.8 & 29.1$\pm$0.7 & 0.195$\pm$0.032 & 72$\pm$18 \\
        & 27 & 18.5$\pm$1.9 & 8.8$\pm$1.2 & 9.3$\pm$0.2 & 27.4$\pm$0.7 & 0.081$\pm$0.008 & 40.9$\pm$7.7 \\
        & 36 & 10.8$\pm$1.1 & 7.9$\pm$1.2 & 9.4$\pm$0.3 & 26.0$\pm$0.6 & 0.157$\pm$0.015 & 50.5$\pm$5.2 \\
        & 37 & 14.3$\pm$1.4 & 8.6$\pm$1.2 & 10.2$\pm$0.2 & 30.4$\pm$1.5 & 0.095$\pm$0.013 & 44.7$\pm$9.8 \\
        & 38 & 19.1$\pm$1.9 & 9.1$\pm$0.8 & 9.7$\pm$0.2 & 34.8$\pm$1.1 & 0.109$\pm$0.009 & 38$\pm$19\\
        & 39 & 11.5$\pm$1.1 & 11.7$\pm$1.8 & 14.4$\pm$0.4 & 34.7$\pm$2.4 & 0.073$\pm$0.018 & 62$\pm$23\\
        & 40 & 19.4$\pm$1.9 & 9.9$\pm$0.6 & 9.9$\pm$0.2 & 25.6$\pm$0.5 & 0.102$\pm$0.009 & 42$\pm$25 \\
        & 49 & 14.6$\pm$1.5 & 9.7$\pm$1.1 & 12.5$\pm$0.3 & 34.3$\pm$1.2 & 0.118$\pm$0.015 & 37.8$\pm$7.6 \\
        & 50 & 14.3$\pm$1.4 & 8.8$\pm$1.2 & 10.6$\pm$0.3 & 34.2$\pm$1.2 & 0.135$\pm$0.014 & 44.4$\pm$7.7 \\
        & 51 & 14.1$\pm$1.4 & 9.7$\pm$1.1 & 13.2$\pm$0.3 & 37.8$\pm$2.0 & 0.073$\pm$0.014 & 47.1$\pm$4.5 \\
        & 52 & 13.5$\pm$1.3 & 9.4$\pm$1.0 & 11.0$\pm$0.3 & 26.7$\pm$1.3 & 0.065$\pm$0.014 & 44.6$\pm$9.4\\
        & 53 & 17.1$\pm$1.7 & 8.6$\pm$1.0 & 9.4$\pm$0.2 & 32.6$\pm$0.8 & 0.116$\pm$0.008 & 53$\pm$28 \\
        & 54 & 25.1$\pm$2.5 & 9.2$\pm$0.7 & 10.1$\pm$0.1 & 30.2$\pm$1.1 & 0.064$\pm$0.008 & 39.7$\pm$6.4 \\
        & 61 & 15.0$\pm$1.5 & 9.3$\pm$1.4 & 11.2$\pm$0.2 & 32.8$\pm$1.6 & 0.064$\pm$0.012 & 36.2$\pm$8.1\\
        & 62 & 13.9$\pm$1.4 & 10.1$\pm$1.4 & 12.4$\pm$0.3 & 37.3$\pm$1.7 & 0.122$\pm$0.015 & 51$\pm$43\\
        & 63 & 13.7$\pm$1.4 & 8.4$\pm$1.1 & 8.9$\pm$0.2 & 26.8$\pm$0.7 & 0.152$\pm$0.013 & 39.1$\pm$9.9 \\
        & 64 & 10.2$\pm$1.0 & 8.4$\pm$1.5 & 11.2$\pm$0.4 & 29.2$\pm$1.0 & 0.131$\pm$0.018 & 50$\pm$13 \\
        & 65 & 13.4$\pm$1.3 & 8.8$\pm$0.8 & 10.0$\pm$0.2 & 22.9$\pm$1.0 & 0.072$\pm$0.014 & 39$\pm$16 \\
        & 72 & 31.7$\pm$3.2 & 11.0$\pm$0.7 & 11.6$\pm$0.1 & 32.1$\pm$1.2 & 0.042$\pm$0.006 & 44.9$\pm$5.4 \\
        & 73 & 21.7$\pm$2.2 & 9.9$\pm$1.1 & 9.7$\pm$0.1 & 33.5$\pm$0.6 & 0.147$\pm$0.008 & 49$\pm$20 \\
        & 74 & 15.4$\pm$1.5 & 8.4$\pm$0.9 & 9.3$\pm$0.2 & 25.3$\pm$0.7 & 0.072$\pm$0.009 & 52$\pm$10 \\
        & 75 & 13.9$\pm$1.4 & 8.8$\pm$1.4 & 10.5$\pm$0.2 & 27.9$\pm$1.4 & 0.084$\pm$0.015 & 37$\pm$15 \\
        & 76 & 18.5$\pm$1.9 & 8.1$\pm$0.7 & 8.6$\pm$0.1 & 26.9$\pm$1.2 & 0.074$\pm$0.010 & 35.6$\pm$3.0 \\
        NGC 4163 & 7 & 6.1$\pm$.61 & 7.3$\pm$1.2 & 9.1$\pm$0.2 & 21.3$\pm$0.6 & 0.06$\pm$0.01 & - \\
        & 8 & 5.5$\pm$.55 & 7.3$\pm$1.5 & 8.8$\pm$0.2 & 21.3$\pm$0.7 & 0.08$\pm$0.01 & 57$\pm$7 \\
        & 15 & 10.2$\pm$1.0 & 8.9$\pm$1.0 & 10.2$\pm$0.1 & 27.4$\pm$0.8 & 0.05$\pm$0.01 & - \\
        & 16 & 10.3$\pm$1.0 & 8.3$\pm$1.2 & 7.4$\pm$0.1 & 20.8$\pm$0.2 & 0.15$\pm$0.01 & 61$\pm$11\\
        & 17 & 13.0$\pm$1.3 & 8.2$\pm$0.8 & 8.0$\pm$0.1 & 21.5$\pm$0.3 & 0.10$\pm$0.01 & 73$\pm$20 \\
        & 24 & 4.4$\pm$.44 & 7.3$\pm$2.0 & 9.4$\pm$0.2 & 21.0$\pm$0.4 & 0.12$\pm$0.01 & 61$\pm$6 \\
        & 25 & 6.4$\pm$.64 & 9.3$\pm$1.6 & 12.6$\pm$0.3 & 40.0$\pm$ 1.2 & 0.01$\pm$0.01 & 60$\pm$4\\
        NGC 6789 & 1 & 17.6$\pm$1.8 & 11.0$\pm$1.2 & 12.3$\pm$0.1 & 19$\pm$12 & 0.005$\pm$0.005 & 38$\pm$7\\
        & 4 & 20.4$\pm$2.0 & 10.2$\pm$0.7 & 12.0$\pm$0.1 & 21.4$\pm$0.2 & 0.031$\pm$0.004 & 41$\pm$6 \\
        UGC 9128 & 2 & 28$\pm$2.9 & 11.0$\pm$1.5 & 14.7$\pm$0.1 & 80$\pm$85 & 0.002$\pm$0.002 & 33.6$\pm$0.8 \\
         & 3 & 15.4$\pm$1.5 & 9.8$\pm$2.6 & 13.3$\pm$0.1 & 29.8$\pm$0.2 & 0.088$\pm$0.003  & - \\
         & 4 & 18.9$\pm$1.9 & 8.4$\pm$2.1 & 9.6$\pm$0.1 & 25.7$\pm$0.1 & 0.186$\pm$0.002 & 35.5$\pm$4.1 \\
         & 5& 17.6$\pm$1.8 & 8.9$\pm$2.1 & 10.5$\pm$0.1 & 25.9$\pm$0.1 & 0.158$\pm$0.002 & 34.3$\pm$2.8 \\[1ex] 
    \enddata
    \tablecomments{A: Regions without H$\alpha$ velocity dispersions were either not covered by the SparsePak observations or no H$\alpha$ flux was detected. }
    \label{vdisps} 
\end{deluxetable*}
\vspace{-20mm}

\subsection{HI Energy Surface Density}

Many studies, including this one, use the velocity dispersion to quantify the turbulence from feedback.  However, the velocity dispersion of a region does not provide the ideal comparison with the SFH.  Due to the differences in column densities between regions, two regions with the same HI energy density may have very different line widths/velocity dispersion.  To account for the HI mass within each region, we measured the HI energy surface density ($\Sigma_{HI}$) along with the velocity dispersion. Between the superprofiles parameters and the second moment averages, three $\Sigma_{HI}$ estimates were used in the analysis:

\begin{enumerate}
    \item $\Sigma_{E,m2}$ is the HI energy surface density from the second moment average derived HI velocity dispersion ($\sigma_{mom2}$)
    \begin{equation}
        \Sigma_{\text{E,m2}} = \frac{3 M_{HI}}{2 A_{HI}}\sigma_{\text{m2}}^2
    \end{equation}
    M$_{HI}$/(A$_{HI}$) is the average HI surface density of the region, where M$_{HI}$ is the HI mass within the region and A$_{HI}$ is the unblotted area of the region.  All regions of a galaxy do not have identical region areas as some regions contain blotted pixels which are removed from the total area of the region, as they do not contribute to the HI mass or velocity dispersion.  The 3/2 factor accounts for the motion in all three directions, assuming isotropic velocity dispersion.
    \item $\Sigma_{E,central}$ is the HI energy surface density derived from the velocity dispersion of the Gaussian fits to the superprofiles ($\sigma_{central}$):
    \begin{equation}
        \Sigma_{\text{E,central}} = \frac{3 M_{HI}}{2 A_{HI}}(1-f_{\text{wing}})(1-f_{\text{cold}})\sigma_{\text{central}}^2
    \end{equation}
    M$_{HI}$ is the total HI mass within the region, \textit{f}$_{\text{wings}}$ is the fraction of HI not in the wings of the superprofile, and (1-f$_{\text{cold}}$) is a correction for the dynamically cold HI ($\sigma < 6$ km s$^{-1}$), which $\sigma_{\text{central}}$ does not describe well M$_{HI}$(1-\textit{f}$_{\text{wings}}$)(1-f$_{\text{cold}}$) is the total HI mass contained within the central peak corrected for the dynamically cold HI, and the fraction of HI within the wings of the superprofile.  We chose $f_{\text{cold}}$=0.15 to be consistent with \cite{Stilp13a} and to be in line with previous estimates for dwarf galaxies \citep{Young03,Bolatto11,Warren12}.  
    \item $\Sigma_{\text{E,wing}}$ is the HI energy surface density derived from the velocity dispersion ($\sigma_{\text{wing}}$) of the wings of the superprofiles:
    \begin{equation}
        \Sigma_{\text{E,wing}} = \frac{3 M_{HI}}{2 A_{HI}}f_{\text{wings}}\sigma_{\text{wings}}^2
    \end{equation}
    M$_{HI}$/(A$_{HI}$)$\times$f$_{\text{wings}}$ represents the total HI surface density associated with the superprofile wings by multiplying the average surface density by the fraction of HI in the wings.
\end{enumerate}
For the HI surface density (M$_{HI}$/(A$_{HI}$)) we assumed 10\% as a reasonable uncertainty based-off the discussion of the accuracy of HI flux measurements and mass determination in \cite{vanzee97}, and accounting for the differences between HI fluxes and masses from single dish observations and the VLA.

\subsubsection{H$\alpha$ Velocity Dispersion ($\sigma_{H\alpha}$)} \label{havdisp}

For each region, we determined which SparsePak fibers fell within the region.  A fiber was placed within a region if more than 50 percent of the area covered by the fiber was within the region.  Due to requiring the detection of H$\alpha$ flux to measure the kinematics of the ionized gas, some regions have ionized gas velocity dispersion measurements based off only a few fibers or do not have ionized gas measurements (see Figures \ref{40683_frame}, \ref{41633_frame}, \ref{67893_frame}, and \ref{91283_frame}).   For each region, the intensity weighted average of the velocity dispersions was measured with: 

\begin{equation}
    \sigma_{H\alpha}=\frac{\Sigma_i \sigma_i F_{H\alpha,i}}{\Sigma_i F_{H\alpha,i}}
\end{equation}
where F$_{H\alpha,i}$ is the H$\alpha$ flux of a fiber, and $\sigma_i$ is the FWHM of the H$\alpha$ line of a fiber.  The average of the $\sigma_{H\alpha}$ was weighted by the line intensity instead of mass, as X-ray observations would be required to determine the mass of the ionized gas. As with the second moment, the standard deviation of a weighted mean was used for the error (see Table \ref{vdisps})

\section{Results: Comparing Star Formation Histories to ISM Turbulence Measures}\label{results}

In this section, we compare the ISM turbulence measures and the SFRs of different time bins from the SFHs to determine over what timescale star formation activity drives turbulence. Determining a strong correlation between the current turbulence and the SFR in a specific time bin would imply that the ISM and star formation activity are coupled on that timescale.

While this analysis is similar to \cite{Stilp13b}, their analysis focused on the global properties of dwarf galaxies. The galactic scale of their analysis of the correlation between HI turbulence and SFH may wash out the impact of star formation activity on smaller scales, as stellar feedback is a local process, on the scale of 10s-100s of parsecs \citep{Kimetal17,Gentry17}.  However, their focus on turbulence on large scales allowed for finer time resolution with step sizes of 10 Myrs in their analysis.  Our focus on the local proprieties of turbulence prohibit similar time resolution.   The differences between the two analyses allow for comparisons between local and global turbulence properties and the timescales involved.

In Section \ref{spearman} we discuss our methods for determining if there is a correlation between the ISM turbulence and SFH, in Sections \ref{Hatimescales} and \ref{HItimescales} we present the results from our initial analysis of four galaxies, and we discuss the implications of the results as well as plans to expand the sample.

\subsection{Spearman Correlation Coefficient} \label{spearman}

To measure the correlation between the ISM turbulence and star formation activity, we used the Spearman rank correlation coefficient $\rho$. The Spearman $\rho$ tests for a monotonic relationship between two variables with a value of $0 < \rho\leq 1$ indicating a positive correlation, a value of $-1\leq \rho < 0$ indicating an anti-correlation, and a $\rho$ of 0 indicates completely uncorrelated data. The value \textit{P} is the probability of finding a $\rho$ value equal to or more extreme than the one measured from a random data set.

\begin{figure}[!th]
    \centering
    \includegraphics[width=.45\textwidth]{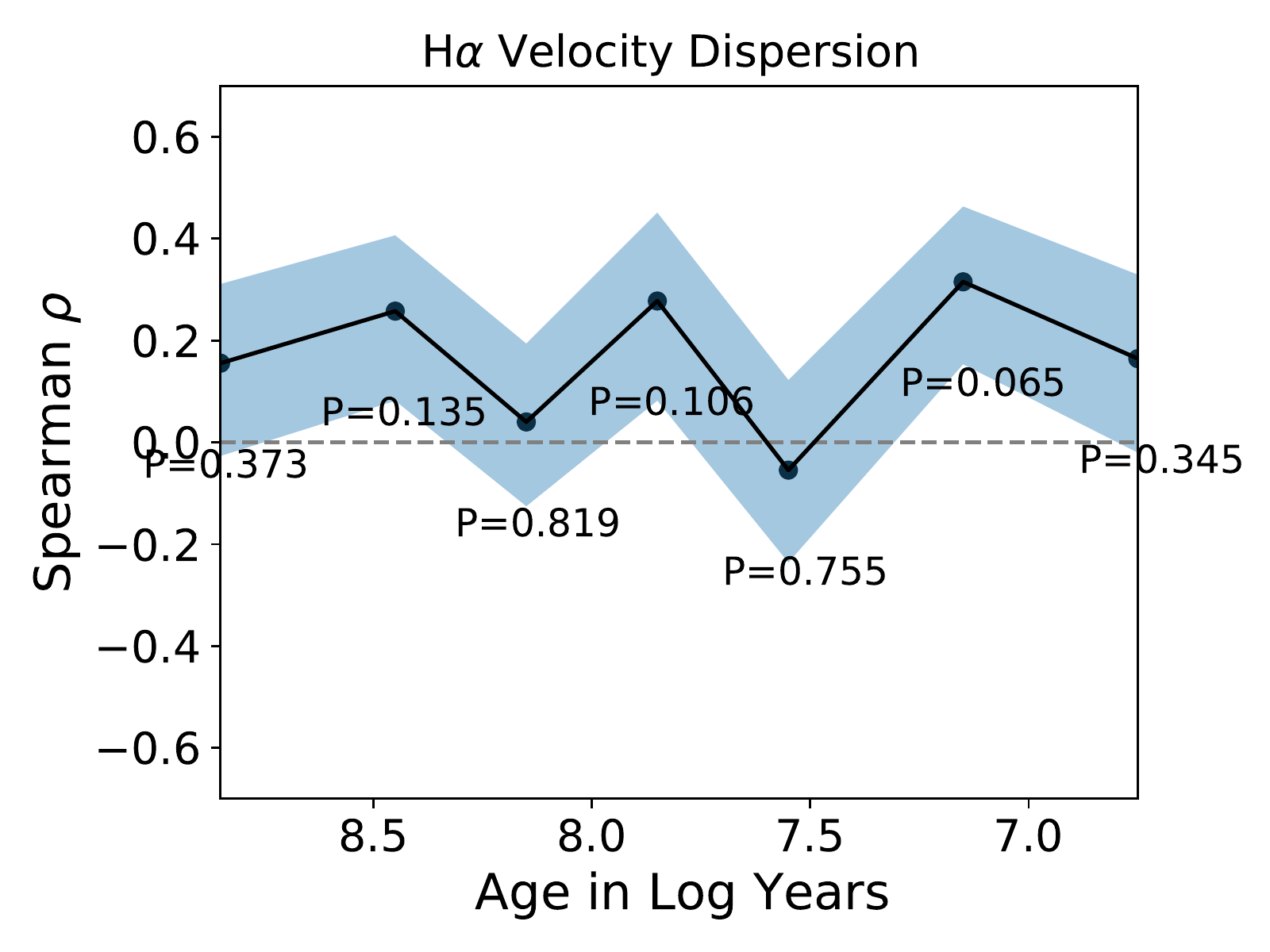}
        \caption{\small The Spearman $\rho$ coefficient versus log time, which demonstrates how correlated the SFR of a given time bin is with the $\sigma_{H\alpha}$. The light blue shaded region represents the 1$\sigma$ bootstrapping error and under each point is the relevant \textit{P} value. For the $\sigma_{H\alpha}$, we do not find a strong correlation between the velocity dispersion and the SFR in any of the time bins.}
    \label{boot_rs_ha}
\end{figure}

\begin{figure*}
    \centering
    \includegraphics[width=.85\textwidth]{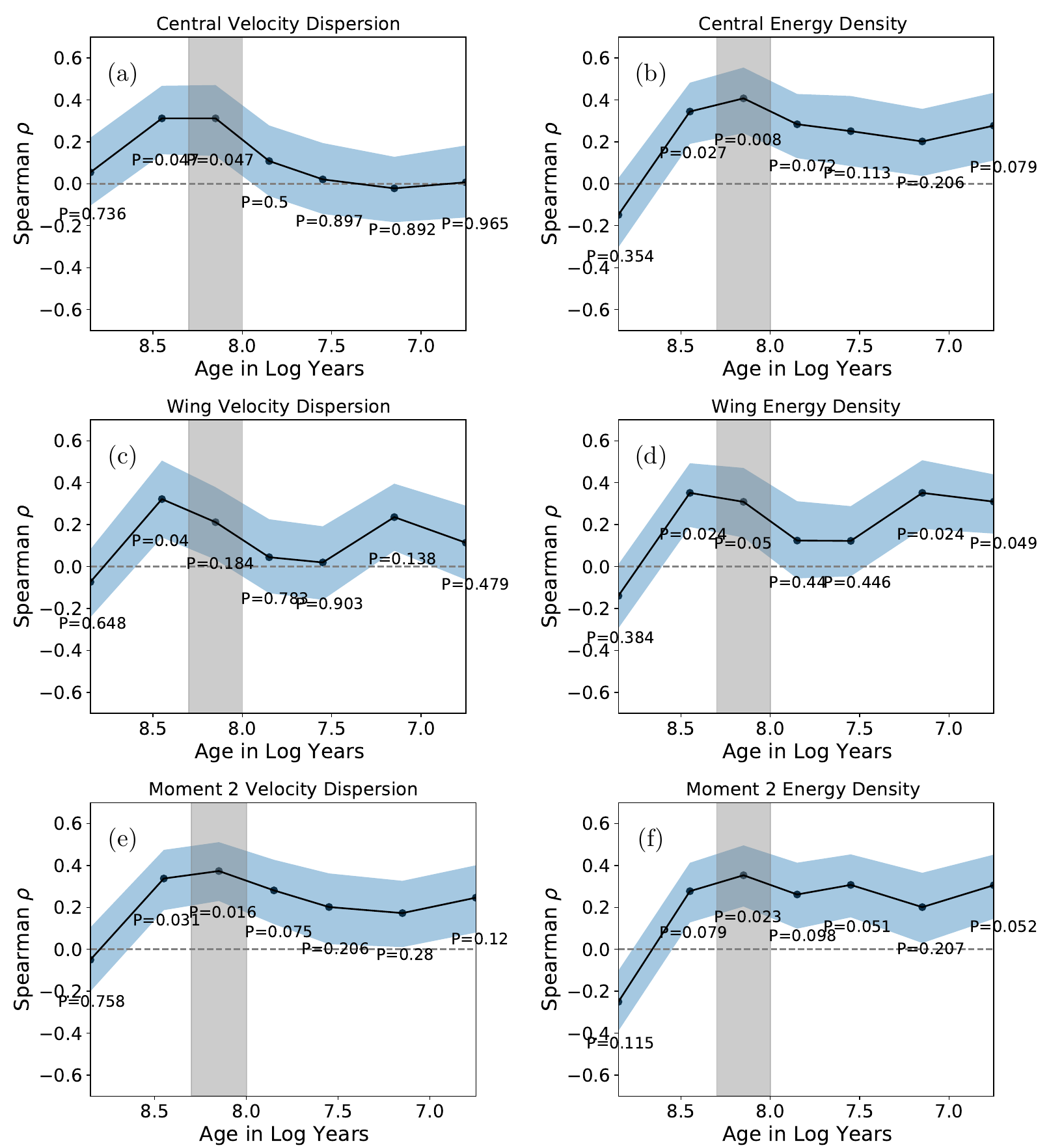}
     \caption{\small The Spearman $\rho$ coefficient and corresponding \textit{P} value plotted against log time showing how correlated the SFR of a given time bin is with the HI turbulence measures. The light blue shaded region represents the 1$\sigma$ bootstrapping error.  a and b) are the correlation of the SFH with the velocity dispersion and energy surface density of the Gaussian superprofiles, c and d) are the correlation of the SFH with the velocity dispersion and energy surface density of the wings of the superprofiles, and e and f) are the correlation of the SFH with the velocity dispersion and energy surface density measured from the second moment maps. For all HI turbulence measures there is at least a modest correlation with the SFR about 100-200 Myrs (8-8.3 log(age)) which is highlighted by the grey shaded region.}
    \label{boot_rs_hi}
\end{figure*}

Each of the seven measures of turbulence (six HI measures, and one H$\alpha$ velocity dispersion), were compared with each SFH time bin.  The resulting $\rho$ and \textit{P} values for the H$\alpha$ correlations are shown in Figure \ref{boot_rs_ha}, and the HI turbulence measures are shown in Figure \ref{boot_rs_hi}. 

To investigate whether or not this selection of regions from 4 galaxies (41 HI regions, 35 H$\alpha$ regions) adequately samples the underlying parameter space, we use bootstrapping to resample the data. We randomly draw a sample of the same size as the existing sample from the original data allowing for repeated values.  Repeating this resampling results in the range of allowable $\rho$ values based on the sample size.  The data were resampled 3000 times and the inner 68$\%$ was taken as the uncertainty on $\rho$.  

\subsection{H$\alpha$ Timescale} \label{Hatimescales}

\begin{figure}[!th]
    \centering
    \includegraphics[width=.45\textwidth]{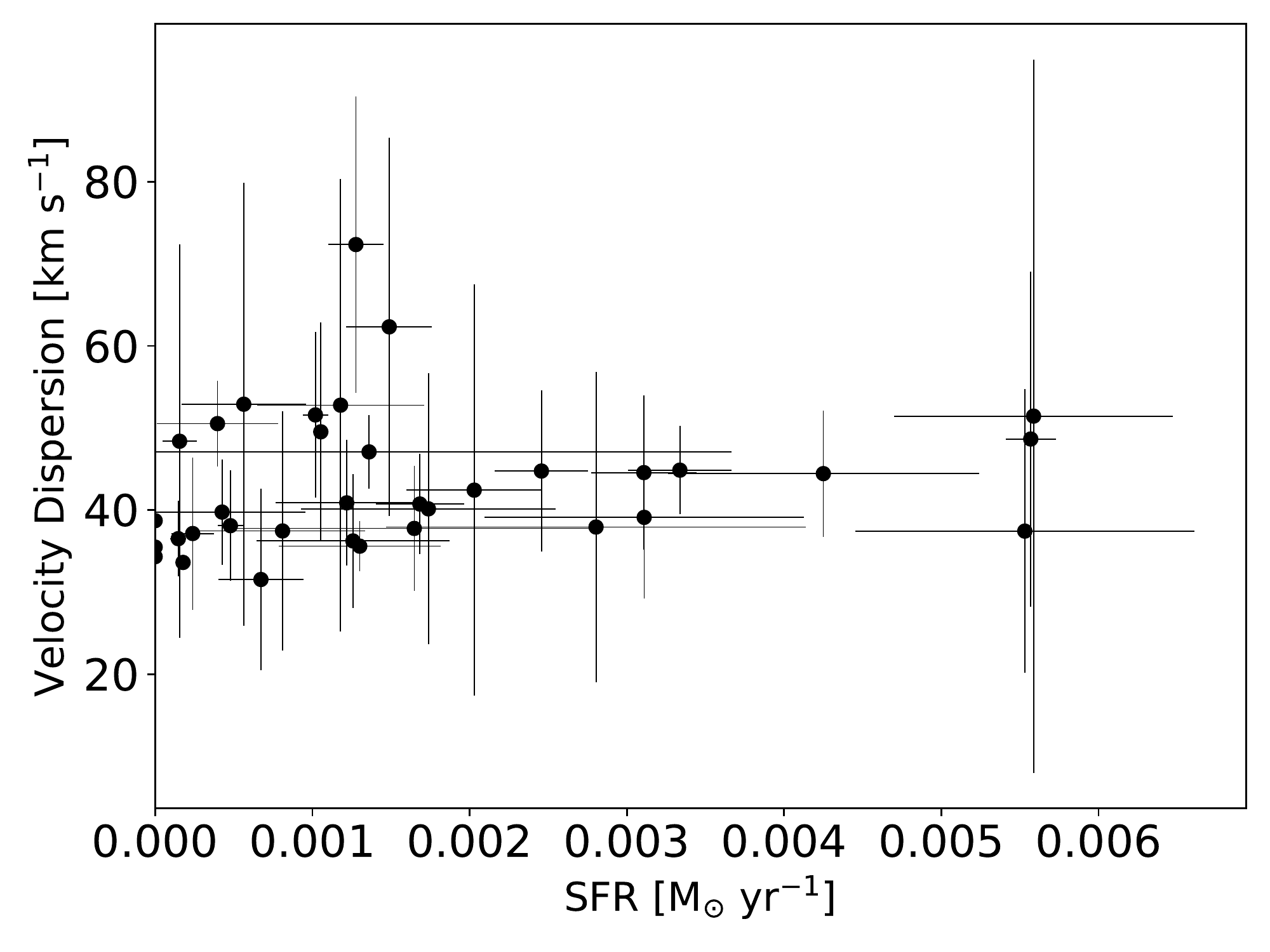}
        \caption{\small The H$\alpha$ velocity dispersion versus the SFR 10-25 Myrs. The error bars represent the 68$\%$ confidence interval of the measurements.}
    \label{sfrvha}
\end{figure}

At this time, our H$_\alpha$ velocity dispersion results are inconclusive. Comparing the $\sigma_{H\alpha}$ and the SFHs, there are no statistically significant correlations. The strongest indication of a correlation seen in Figure \ref{boot_rs_ha} is between the $\sigma_{H\alpha}$ and star formation activity 10-25 Myrs ago, however, as the uncertainties on the $\sigma_{H\alpha}$ are significant, any trend of higher velocity dispersion at higher SFR is overwhelmed by the uncertainties in Figure \ref{sfrvha}.  Over all in Figure \ref{boot_rs_ha} there is the suggestion of a positive correlation between the H$\alpha$ velocity dispersion and the cumulative SFH. Such a correlation would indicate that the ionized gas turbulence is related to the star formation history, as is expected if stellar feedback drives turbulence.  These four galaxies were a test of the methods described in this paper and represent a small subset of a larger sample. It is possible that an increased number of galaxies and regions will allow for the selection of regions with reliable $\sigma_{H\alpha}$ to constrain the timescale over which stellar feedback drives turbulence in the ionized gas. 

As the SFHs do not provide SFR for the past 5 Myrs (see \cite{McQuinn10a} for details), we are not sensitive to a correlation between the $\sigma_{H\alpha}$ and the current SFR which has been observed in previous IFU analysis comparing H$\alpha$ derived SFRs and $\sigma_{H\alpha}$ \citep{Moiseev15,Zhou17}. To analyse the correlation between the ionized gas turbulence and the SFR over the past 5 Myrs for individual regions requires sufficient H$\alpha$ flux within each region. For the majority of regions within the four galaxies used for this paper, H$\alpha$ derived SFRs would be highly uncertain due to the low H$\alpha$ fluxes.  With a larger sample, it may be possible to have sufficient regions to accurately measure SFRs over the past 5 Myrs and compare with the ionized gas turbulence.

\subsection{HI Timescale} \label{HItimescales}

\begin{figure}[!th]
    \centering
    \includegraphics[width=.45\textwidth]{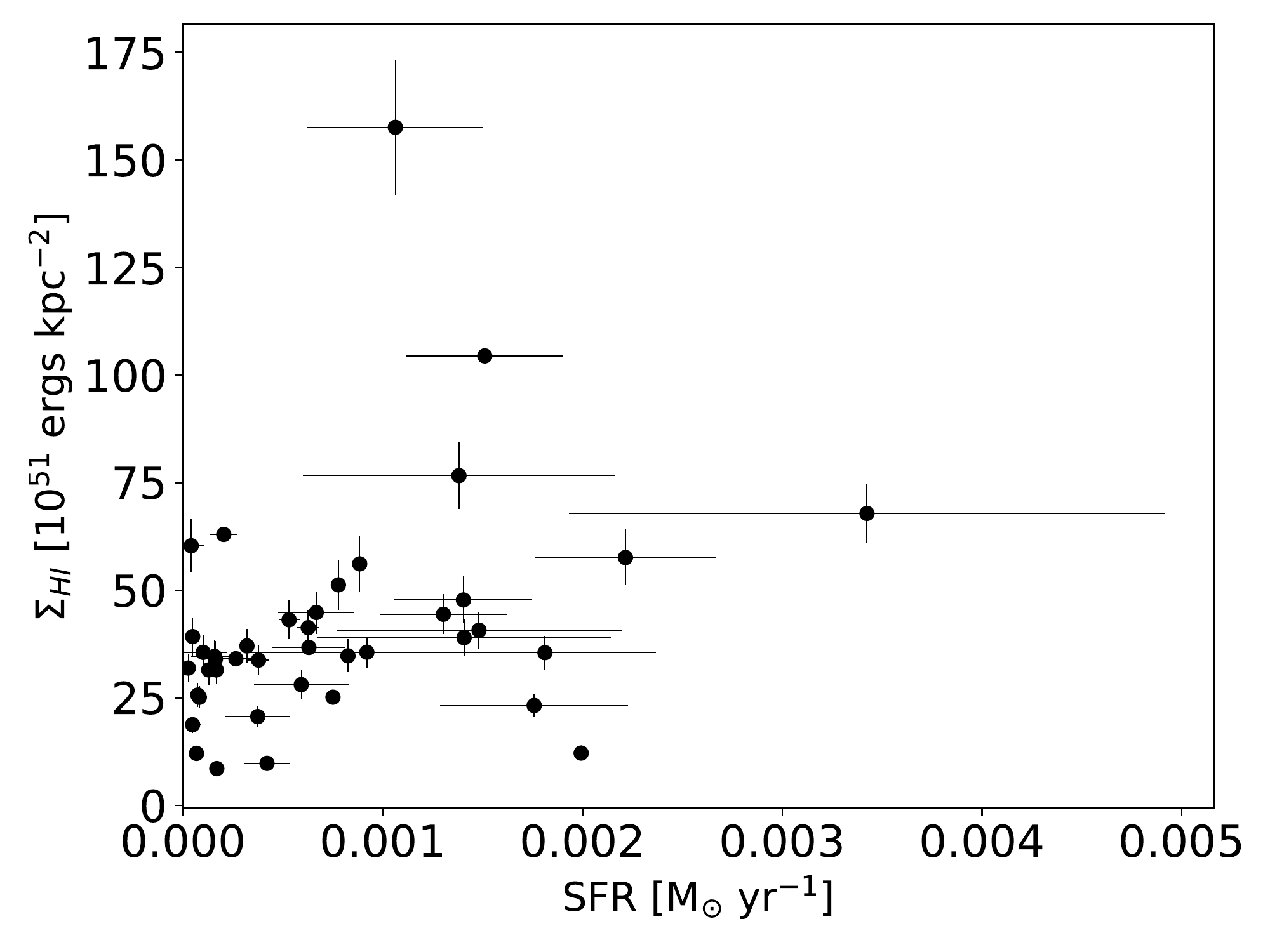}
        \caption{\small The HI energy surface density ($\Sigma_{HI}$) of the Gaussian superprofile fit versus the SFR 100-200 Myrs.  The error bars on the SFRs represent the 68$\%$ confidence interval of the measurements. A slight trend of higher $\Sigma_{HI}$ with higher SFR can be seen. }
    \label{sfrvhi}
\end{figure}

In Figure \ref{boot_rs_hi} there is a modest peak in the Spearman $\rho$ value when comparing multiple measures of the current HI turbulence and the SFR 100-200 Myrs ago.  This modest correlation can be seen when comparing the HI velocity dispersions and $\Sigma_{HI}$ measured from the scaled-Gaussian fit (subfigures \ref{boot_rs_hi}a and \ref{boot_rs_hi}b) and from the second moment maps (subfigures \ref{boot_rs_hi}e and \ref{boot_rs_hi}f). The strongest correlation between the HI turbulence and past star formation activity is between the energy surface density of the superprofiles and the SFR 100-200 Myrs ago (Figure \ref{boot_rs_hi}b).  The measured $\rho$ is 0.407 and the \textit{P} value is 0.008.  The SFR and the $\Sigma_{HI}$ can be seen in Figure \ref{sfrvhi}, where the correlation is dominated by the handful of regions with high SFRs or high $\Sigma_{HI}$.  

The correlation observed between the HI turbulence and SFH 100-200 Myrs may be related to the time scales over which turbulent momentum decays.  \cite{Bacchini20a} found for SNe the estimated dissipation time for the atomic gas ranges between a few 10s of Myrs and 100s of Myrs depending on the disk thickness. Similarly, from FIRE-2 simulations, \cite{Orr2020} theorized that the strong correlation between the ISM turbulence and the SFR over 100 Myrs in the simulation may be because 100 Myrs is approximately the eddy-crossing time. As a result of long dissipation times for turbulent momentum, the velocity dispersion may evolve slowly and the impact of older star formation activity could remain observable in the ISM. 

Due to their ability to trace back star formation activity and determine ISM turbulence on the relevant scales, simulations of dwarf galaxy evolution provide on excellent comparison to results presented here.  Whether or not the same timescale is observed in simulations would be of interest. The timescales observed in simulations could either support of the results presented here, or open new questions about the implementation of feedback and turbulence in dwarf galaxies and the handling of the atomic gas in simulations.

Previously, \cite{Stilp13b} found the strongest correlation between the HI energy surface density and the SFR 30-40 Myrs ago in their study of 18 galaxies.  However,  \cite{Stilp13b} traced the SFH of their galaxies back only 100 Myrs, and as such were not sensitive to the correlation timescale we measured.  In Figure \ref{boot_rs_hi} no evidence of a correlation between the HI turbulence and the SFRs measured in the 25-50 Myr time bin is seen, where we would expect to see a correlation based on \cite{Stilp13b}. Because of their focus on global properties, \cite{Stilp13b} used 10 Myr time bins for their study and so are more sensitive to the impact of short timescale variations in the SFH compared to our analysis where the finer spatial resolution prevents finer time resolution.  The difference in time binning may decrease the amplitude of the correlation between the HI turbulence and the SFR in the relevant time bin.  The difference in the observed correlation timescales could indicate a difference between the global and local turbulence properties of galaxies, and that the impact of stellar feedback on the ISM is scale dependent. By analyzing turbulence on different physical scales, a more complete picture of the interplay between stellar feedback and turbulence is made.   

\subsubsection{The Influence of NGC 4068 on These Results}

Three of the four galaxies analyzed in this paper are physically small and, as such, have small numbers of regions.  The fourth galaxy, NGC 4068, is significantly larger than the other three and contains over half the regions analyzed for this paper.  NGC 4068's inclusion in the initial sample is important as it greatly increases our region sample size and our ability to test our methods and draw preliminary conclusions from a small sample of galaxies.  However, we must consider if NGC 4068's large number of regions are dominating the results. 

For the H$\alpha$ velocity dispersion, we repeated the analysis excluding NGC 4068 from the sample which results in 10 regions with SparsePak measurements, half of which have very sparse coverage. Analyzing the regions in NGC 4163, NGC 6789, and UGC 9128 results in no correlation between the $\sigma_{H\alpha}$ and the SFH in the past 500 Myrs, same as when the full sample of regions was analyzed. 

\begin{figure*}[!ht]
    \centering
    \includegraphics[width=.85\textwidth]{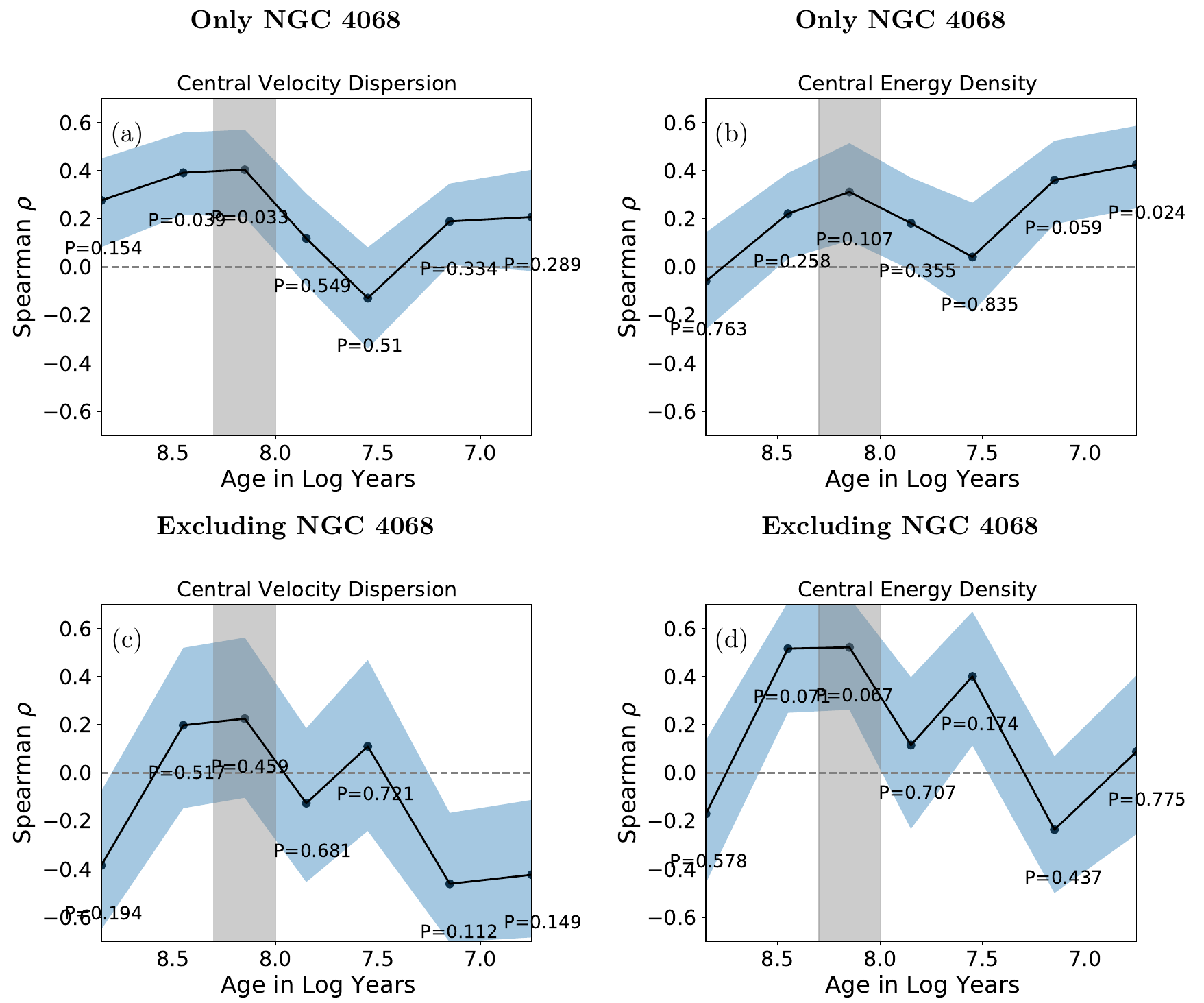}
     \caption{\small The Spearman $\rho$ coefficient and corresponding \textit{P} value plotted against log time showing how correlated the SFR of a given time bin is with the velocity dispersion and energy surface density of the Gaussian superprofiles. The light blue shaded region represents the 1$\sigma$ bootstrapping error.  A and B) are the correlations for regions within NGC 4068 C and D) are the correlation for the regions within NGC 4163, NGC 6789, and UGC 9128. For all HI turbulence measures, there is a peak in the correlation at about 100-200 Myrs.}
    \label{boot_rs_hi_cen}
\end{figure*}

For the HI turbulence, we repeated the analysis twice, once excluding the regions in NGC 4068, and once only analyzing regions in NGC 4068.  Both data sets have indications of the 100-200 Myr correlation timescale and demonstrates the results are not dependent on the inclusion of NGC 4068 in the sample. The correlation between the SFH 100-200 Myr ago and the HI turbulence is not as prominent when including only the three smaller galaxies, or only studying NGC 4068, compared to when the entire sample is analyzed.  In Figure \ref{boot_rs_hi_cen} there are peaks in $\rho$ at 100-200 Myrs ago in subfigures A, C and D. For NGC 4068, there is a statistically significant correlation between the velocity dispersion and the SFR 100-200 Myrs.  However, there is no significant correlation with the $\Sigma_{HI}$.  For the three smaller galaxies, the correlation between the SFR 100-200 Myrs ago and the velocity dispersion is non-existent, while the correlation with the $\Sigma_{HI}$ is not statistically significant with a \textit{p} value of 0.067.  Similar results are seen for the other measures of HI turbulence when analyzing the two subsamples.  There are clear peaks in $\rho$ at the 100-200 Myr timescale, but the peaks are rarely statistically significant.  The inclusion of NGC 4068 is not dominating the results of the HI, as all four galaxies are responsible for the correlation seen in Figure \ref{boot_rs_hi}.

\section{Summary} \label{conclusion}

In this paper, we outlined our methods for determining the timescales over which star formation drives turbulence in the ISM on a spatially resolved scale of $\sim$400 pc. We described how we analyzed available \textit{HST}, VLA, and SparsePak (WIYN 3.5m) observations of the four galaxies (NGC 4068, NGC 4163, NGC 6789, and UGC 9128) included in the initial study.  Using these four galaxies as examples, we detailed how we selected the regions of interest and how we measured the SFH, HI turbulence and H$_\alpha$ velocity dispersion in each region.

With this initial sample, we compared the local HI energy surface density ($\Sigma_{HI}$), measured from Gaussian superprofiles and second moment maps and $\sigma_{H\alpha}$ with spatially resolved SFHs. Using the Spearman's rank correlation coefficient, we found the strongest correlations between the SFH and the atomic gas velocity dispersion and energy surface density are seen between 100 and 200 Myrs ago. 

A strong correlation between the HI turbulence measures with the SFR 100-200 Myr was unexpected and may be related to the time scales over which turbulent momentum decays.  This correlation may be due the dissipation times of dwarf galaxies which is on the scale of 100 Myrs. Long dissipation times for turbulent momentum would result in the velocity dispersion evolving slowly and the impact of older star formation activity may remain observable in the ISM.  

With the selection of four galaxies, we are limited in our ability to draw broad conclusions and are left asking: is the measured 100-200 Myrs timescale universal or does the timescale vary based on a galaxy's characteristics?  Differences in the physical properties of galaxies could result in a varying correlation timescale between stellar feedback and turbulence.   The four galaxies included in this paper are all members of STARBIRDS \citep{McQuinn15a}, and are currently or recently starbursting galaxies. This common feature in the galaxies' SFHs may impact the timescales involved compared to less active recent SFHs.  A more diverse sample of galaxies will help assess if a galaxy's physical characteristics play an essential role in how stellar feedback and the ISM are connected.

As previously mentioned, these four galaxies were a test of the methods described in this paper and represent a small subset of a larger sample. The total planned sample includes low-mass (log(M$_{odot}$)=6-9.5) star forming galaxies within 5 Mpc with a range of current SFRs and SFHs. The planned sample, with its larger range of galaxy characteristics will permit the analysis of how certain galactic properties may alter the 100-200 Myrs correlation timescale. Along with analyzing how recent SFH impacts the correlation timescale, another key galaxy property that may impact the correlation timescale is metallicity. Variations in metallicity cause variations in the cooling timescale of the ISM as thermal energy dissipates at different rates. Such differences in cooling timescale may impact the observed correlation timescale between star formation activity and turbulence in the ISM. A broader selection of galaxies allows for the grouping of galaxies with similar characteristics to probe the importance of parameters such as mass, metallicity, and recent SFH on the correlation timescale. This initial study sets the framework for a larger investigation of feedback and turbulence in low-mass galaxies

\begin{center}
    \textbf{ACKNOWLEDGEMENTS}
\end{center}

\begin{acknowledgments}
    The authors would like to thank Justin A. Kader for his assistance in visualizing the ionized gas data and Emily E. Richards for her early work on NGC 4068 and NGC 6789. This work is financially supported through NSF Grant Nos. AST-1806926 and AST-1806522. Any opinions, findings, and conclusions or recommendations expressed in this material are those of the authors and do not necessarily reflect the views of the National Science Foundation. Work in this paper was partially supported by NSF REU grant PHY-1460882.  The authors acknowledge the observational and technical support from the National Radio Astronomy Observatory (NRAO), and from Kitt Peak National Observatory (KPNO). Observations reported here were obtained with WIYN 3.5 telescope which is a joint partnership of the NSF's National Optical-Infrared Astronomy Research Laboratory, Indiana University, the University of Wisconsin-Madison, Pennsylvania State University, the University of Missouri, the University of California-Irvine, and Purdue University.  This research made use of the NASA Astrophysics Data System Bibliographic Services and the NASA/IPAC Extragalactic Database (NED), which is operated by the Jet Propulsion Laboratory, California Institute of Technology, under contract with the National Aeronautics and Space Administration.
    
    \textit{Facilities:} Hubble Space Telescope; the Very Large Array; the WIYN Observatory\\
    \textit{Software:} Astropy \citep{astropy13,astropy18}; GIPSY \citep{GIPSY}; Peak ANalysis \citep{PAN}; IRAF \citep{IRAF86,IRAF93}
\end{acknowledgments}
 
{\small\bibliography{ref}}

\end{document}